\documentclass[12pt]{article}

\usepackage{latexsym,amsmath,amssymb,theorem,epsfig}
\usepackage{graphicx}

\DeclareFontFamily{OT1}{rsfs10}{}
\DeclareFontShape{OT1}{rsfs10}{m}{n}{ <-> rsfs10 }{}
\DeclareMathAlphabet{\mathscript}{OT1}{rsfs10}{m}{n}

\numberwithin{equation}{section}

\newcommand{\ns}{\normalsize}

\def\cO{{\mathcal O}}

\def\cV{{\mathcal V}}

\def\gsim{ \lower .75ex \hbox{$\sim$} \llap{\raise .27ex \hbox{$>$}} }
\def\lsim{ \lower .75ex \hbox{$\sim$} \llap{\raise .27ex \hbox{$<$}} }
\def\be{\begin{equation}}
\def\ee{\end{equation}}
\def\bea{\begin{eqnarray}}
\def\eea{\end{eqnarray}}

\theoremstyle{plain}

{\theorembodyfont{\rmfamily} }

\begin{document}


\begin{titlepage}

\title{
  \hfill{\ns }  \\
   {\LARGE On the Initial Conditions in New Ekpyrotic Cosmology}
\\}
\author{
   Evgeny I. Buchbinder$^{1}$, Justin Khoury$^{1}$, Burt A. Ovrut$^{2}$
     \\[0.5em]
   {\ns ${}^1$ Perimeter Institute for Theoretical Physics} \\[-0.4cm]
{\ns Waterloo, Ontario, N2L 2Y5, Canada}\\[0.3cm]
{\ns ${}^2$ Department of Physics} \\[-0.4cm]
{\ns The University of Pennsylvania}\\[-0.4cm]
{\ns  Philadelphia, PA 19104--6395, USA}\\[0.2cm]}

\date{}

\maketitle

\begin{abstract}

New Ekpyrotic Cosmology is an alternative scenario of early universe cosmology in which the universe
existed before the big bang. The simplest model relies on two scalar fields, whose entropy perturbation leads to a scale-invariant spectrum of density fluctuations.
The ekpyrotic solution has a tachyonic instability along the entropy field direction which, a priori, appears to require fine-tuning of the initial conditions. 
In this paper, we show that these can be achieved naturally by adding a small positive mass term for the tachyonic field and coupling to light fermions. 
Then, for a wide range of initial conditions, the tachyonic field  gets stabilized with the appropriate values 
well before the onset of the ekpyrotic phase. Furthermore, we show that ekpyrotic theory is successful in solving the flatness, horizon and homogeneity problems of standard big bang
cosmology. Motivated by the analysis of the tachyonic instability, we propose a simplification of the model in terms of new field variables. Instead of requiring two
exponential potentials, one for each scalar field, it suffices to consider a single nearly exponential potential for one of the fields and a tachyonic mass term along the orthogonal direction in field space. 
All other terms in the potential are essentially arbitrary. This greatly widens the class of ekpyrotic potentials and allows substantial freedom in determining the  spectral index and possible non-Gaussianity. We present a generalized expression for the spectral index which is easily consistent with the observed red tilt. We also argue that for a wide range of potentials non-Gaussianity can be substantial, within the reach of current observations.

\end{abstract}

\thispagestyle{empty}

\end{titlepage}

\section{Introduction}

The ekpyrotic scenario~\cite{ek1,seiberg,ekpert,briefing} is an alternative theory to the standard inflationary big bang paradigm. Instead of invoking a flash of exponential
expansion shortly after the big bang, ekpyrotic theory proposes that the universe started out in a cold state followed by a long period of slow
contraction. Despite this stark difference in dynamics, remarkably the two models make identical predictions for the origin of structure formation:
a nearly scale-invariant, adiabatic and Gaussian spectrum of density perturbations. One testable distinctive prediction is that inflation also generates scale-invariant primordial gravitational waves whose amplitude in the simplest models is within reach of near-future microwave background polarization experiments~\cite{compare}, whereas the ekpyrotic scenario does not~\cite{ek1,gwaves}. 
The idea of a pre-big bang contracting phase arising from nearly vacuous initial conditions originated in the pre-big bang scenario of Gasperini
and Veneziano~\cite{pbb}. 

Until recently, ekpyrotic theory suffered from two important drawbacks. 

First, it was unknown how to describe a non-singular bounce within a consistent effective theory without introducing ghosts or other catastrophic
instabilities~\cite{instabilities}. This lead the advocates of the Big Bang/Big Crunch ekpyrotic~\cite{seiberg, ekpert} and cyclic models~\cite{cyc,design} to propose that the universe
undergoes a big crunch singularity to be resolved by stringy physics. This seems plausible in light of the mildness of the singularity. In the context of colliding branes in heterotic M-theory~\cite{het1,het1A}, for instance, the singularity corresponds to the fifth dimension shrinking to zero size, with the large three spatial dimensions remaining finite. Despite intense activity in recent years~\cite{bunch}, however, a definitive example of a cosmological bounce in string theory is still missing.

Second, the fate of perturbations through the bounce, and therefore the prediction of scale invariance, is ambiguous. This stems from the fact that
the scale-invariant growing mode of the scalar field fluctuations precisely projects out of $\zeta$~\cite{ekpert,condscaleinv} --- the curvature perturbation on uniform-density
hypersurfaces~\cite{bardeen}. The latter is a useful variable to track since it is conserved on super-horizon scales. For this reason, many have argued that
the outcome of the bounce would amount to matching $\zeta$ continuously, resulting in an unacceptably blue spectral tilt~\cite{zetanay}. 
While this is universally true for any non-singular bounce within 4d Einstein gravity, it is nevertheless conceivable that stringy or higher-dimensional effects relevant near the bounce
may lead to mode-mixing and impart $\zeta$ with a scale-invariant piece~\cite{zetayay}.

In a recent paper~\cite{us}, we proposed a fully consistent and complete scenario which addresses both issues. While sharing many important ingredients of the
old ekpyrotic scenario~\cite{ek1,ekpert,negtension}, this model of New Ekpyrotic Cosmology generates a non-singular bounce, all describable within a 4d effective field theory, and unambiguously leads to a scale-invariant curvature perturbation. 

The physics of the bounce exploits the mechanism of ghost condensation~\cite{Arkani}, which relies on higher-derivative corrections to the kinetic term for a scalar field, akin to
k-inflation~\cite{kinf} or k-essence~\cite{kess}. Ghost condensation allows for violations of the null-energy condition (NEC) without introducing ghosts or other pathologies~\cite{Paolo}. In~\cite{us} we showed explicitly how the ghost condensate can be merged consistently with the preceding ekpyrotic phase to generate a smooth transition from contraction to expansion.
Whether ghost condensation can be realized in a UV complete theory, such as string theory, remains an open issue~\cite{IRnima}, although see~\cite{Mukohyama} for a recent attempt.
Even if a smooth bounce is realized by very different physics, many of our key results will still apply. For our purposes the ghost condensate provides an explicit and tractable realization of such a bounce.

In our model $\zeta$ acquires a scale-invariant spectrum well-before the bounce through the conversion of entropy (or isocurvature) perturbations. By
having two scalar fields, each with their own ekpyrotic potentials, the entropy perturbation --- corresponding to the
difference in the scalar field fluctuations --- is scale-invariant~\cite{us,private,paologhost,finelli1}. See also~\cite{notari,finelli2,finelli3}. 
In~\cite{us}, we showed how the entropy mode gets converted into the adiabatic mode by using features in the potential
which are independently necessary for ending the ekpyrotic phase and bridging into the bounce. Then, since the physics of
the bounce is all within 4d effective theory, $\zeta$ goes through the bounce unscathed and emerges in the hot big bang phase
with a nearly scale-invariant spectrum.

As pointed out in~\cite{us,private} and studied extensively in~\cite{wands1,wands2}, the model has a tachyonic instability along the entropy direction. Since this instability is at the origin of the growth of entropy modes, the tachyon is unavoidable~\cite{tolley}. As it stands, New Ekpyrotic Cosmology
would ostensibly require fine-tuned initial conditions in order for the field trajectory to start out rolling along this tachyonic ridge.
In this paper, we show how such initial conditions can be achieved naturally with a pre-ekpyrotic stabilization phase.   

Our original scenario~\cite{us} was cast in terms of two scalar fields, $\phi_1$ and $\phi_2$, each with their own steep, negative exponential potential. Through a rotation in field space, following~\cite{wands1} we can instead study the dynamics in terms of new field variables $\phi$ and $\chi$, respectively the field directions along and orthogonal to the field trajectory.
This is reviewed in Sec.~\ref{2field}. In this language, the solution describes rolling down an ekpyrotic potential along $\phi$, while $\chi$ remains fixed at a tachyonic point, thereby making the instability manifest.

The $(\phi,\chi)$ perspective leads us to propose a significant simplification and generalization of the scenario, described in Sec.~\ref{simplified}. By treating $\phi$ and $\chi$ as fundamental fields, 
the origin of the potential in terms of the {\it two} steep and quasi-exponential potentials becomes unnecessary. All we need is a {\it single} steep and nearly exponential potential along
$\phi$, as well as a tachyonic mass for $\chi$, leaving tremendous freedom in specifying the global shape of the potential. The only constraint which is crucial in generating
a scale-invariant spectrum is that the curvature of the potential along the $\chi$ and $\phi$ directions must be nearly the same: $V_{,\phi\phi}\approx V_{,\chi\chi}$.

\begin{figure}[htb]
\begin{center}
\includegraphics[width=120mm]{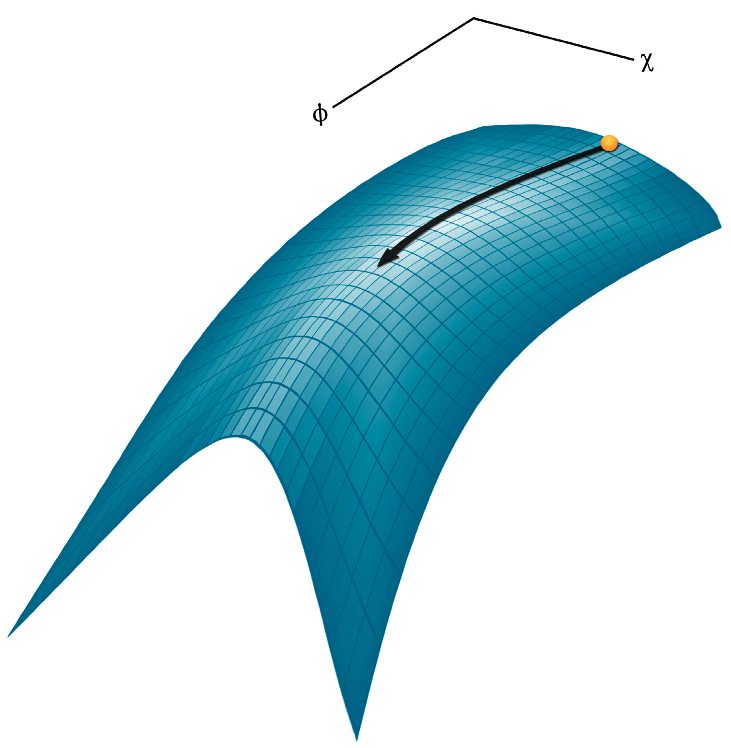}
\caption{General shape of the potential during the ekpyrotic phase. The arrow indicates the desired solution, corresponding to rolling down a steep, negative and quasi-exponential potential along $\phi$, while remaining perched on top of a tachyonic ridge along $\chi$.}
\end{center}
\label{2fieldpotek}
\end{figure}

This novel framework greatly expands the class of allowed ekpyrotic potential --- see Fig.~\ref{2fieldpotek} for a generic example.
In Sec.~\ref{tilt} we derive the spectral tilt for the most general New Ekpyrotic potential and find that it depends on three parameters:
the usual fast-roll parameters $\epsilon$ and $\eta$ characterizing respectively the steepness and deviation from pure exponential
form of the potential, as well as a new parameter $\delta$ describing the difference in curvature between the $\chi$ and $\phi$
directions. The class of potentials studied in~\cite{us} corresponds to $\delta=0$. Here we see that turning on $\delta$ allows for even greater freedom in the spectral
tilt. In particular, pure exponential potentials ($\eta = 0$) were found to give a blue spectrum in~\cite{us,private,paologhost}, in disagreement with the recent WMAP data~\cite{wmap3}.
This led~\cite{us,private} to consider deviations from the pure exponential form ($\eta\neq 0$), which does allow for a small red tilt.
Here, however, we see that even for pure exponentials the tilt can be red by having non-vanishing $\delta$. 

In Sec.~\ref{nongauss}, we discuss the generic form of
higher-order terms in $\chi$. These self-interaction terms play a crucial role in determining the level of non-Gaussianity of the perturbation spectrum.
We find that the non-Gaussian signal generated during the ekpyrotic phase is generically substantial, within the reach of current and near-future experiments.

The tachyonic instability in the $\chi$ direction is part of a more general discussion of initial conditions in New Ekpyrotic Cosmology, in particular the assumed degree of initial flatness and homogeneity.
In the context of the original ekpyrotic scenario~\cite{ek1} it was initially believed that the model did not address the homogeneity and flatness problems of standard big bang cosmology~\cite{pyro}. See~\cite{lindenew} for a recent critique of the scenario along these lines. Such criticism motivated the advent of the cyclic model~\cite{cyc,design}, where homogeneity and flatness in a given cycle is achieved through a phase of dark energy domination in the previous cycle. It was recently realized, however, that the ekpyrotic model does in fact solve these problems~\cite{cyclicages}, in a way akin to inflationary cosmology. 

In Sec.~\ref{flatness}, we argue that New Ekpyrotic Cosmology addresses the flatness, homogeneity
and isotropy problems of standard big bang cosmology without invoking any phase of cosmic
acceleration. This relies crucially on the ekpyrotic attractor mechanism --- during the ekpyrotic phase, the energy density in the
ekpyrotic scalar field blueshifts much faster than any other component, in particular spatial curvature and anisotropic
stress. In other words, contrary to the naive expectation that a contracting phase makes the universe more inhomogeneous,
here the universe becomes increasingly smooth.
Therefore, making the natural assumptions that the proto-ekpyrotic patch is fairly homogeneous, isotropic, and has curvature comparable to the Hubble scale,
then, provided the ekpyrotic phase lasts for sufficient number of e-folds, the universe emerges in the hot big bang phase with
the high level of symmetry we observe today. As such, New Ekpyrotic Cosmology resolves the standard problems of the big bang model.

This is precisely analogous to inflationary cosmology. There one assumes that over some macroscopic patch the universe is sufficiently
homogeneous and flat to allow inflation to take over. Subsequently, cosmic acceleration flattens and homogeneizes the universe to
the high level we observe it today. Of course, in absolute terms the assumed initial radius of curvature is much larger in ekpyrosis than in inflation, but by no means
is this a drawback of one model versus the other. Instead, this corresponds to drastically different assumptions about the initial state of the universe: 
inflation assumes a hot initial state, with correspondingly high Hubble scale; ekpyrosis proposes a cold beginning, with correspondingly small expansion rate. 
In the absence of a concrete and rigorous theory of initial conditions, either starting point is equally reasonable. With respect to flatness and homogeneity, the proper yardstick is, therefore,
the assumed level of flatness and homogeneity compared to the respective initial Hubble radius. And, from this point of view, both models are equally successful.

\begin{figure}[htb]
\centering
\includegraphics[width=120mm]{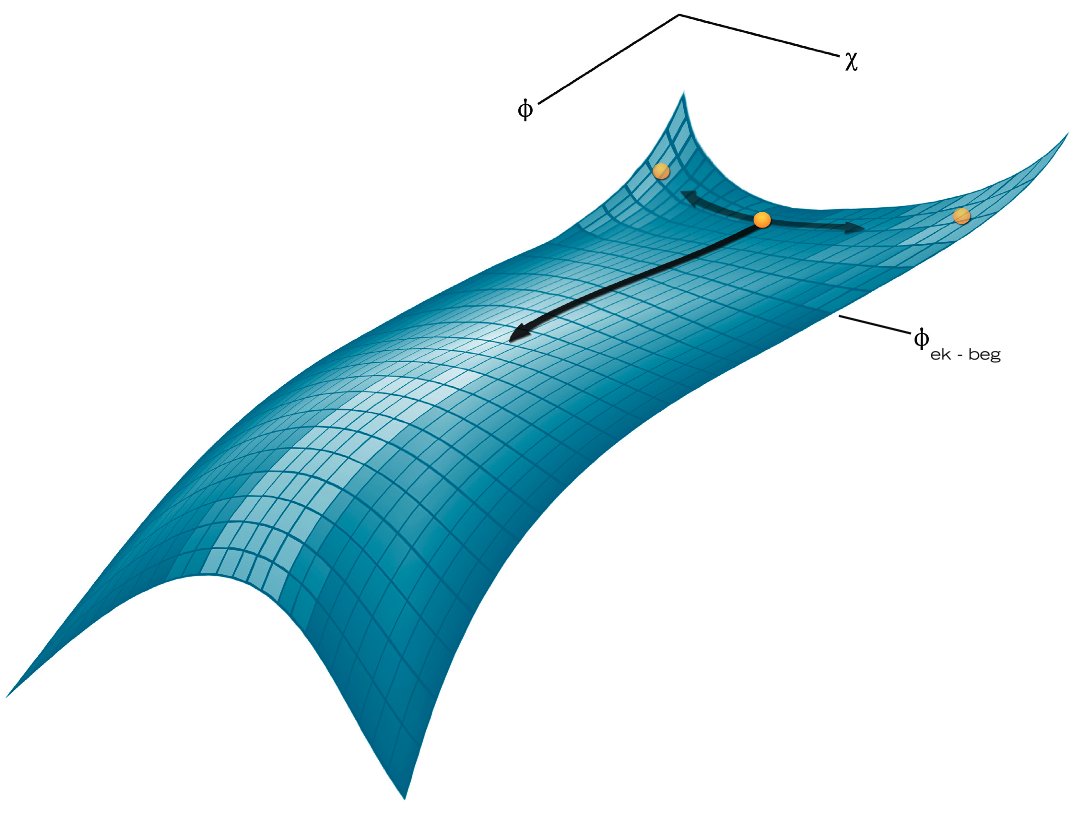}
\caption{A pre-ekpyrotic mechanism sets up the desired initial conditions for the ekpyrotic phase. With the addition of a positive mass-squared term, the field is initially stable in the $\chi$ direction. By introducing couplings to light fermions, the field settles down to the minimum for a wide range of initial conditions by the onset of the ekpyrotic phase.}
\label{2fieldpotstab}
\end{figure}

In this paper, we address in a simple and natural way the fine-tuning of initial conditions associated with the aforementioned tachyonic instability of the two-field model.
In Sec.~\ref{taming} we quantify the nature of the instability and discover that exponentially fine-tuned conditions 
must be satisfied at the onset of the ekpyrotic phase in order for the roll along the tachyonic ridge to last sufficiently long to produce enough e-foldings of scale-invariant perturbations. 
We then solve this initial fine-tuning by introducing a positive mass squared term for $\chi$, which is relevant and stabilizes $\chi$ at early, pre-ekpyrotic times, but becomes negligible
during the ekpyrotic phase. In other words, this term serves to set up the desired initial conditions but does not jeopardize the subsequent  generation of perturbations. Thus, $\chi$ starts oscillating around its stable minimum. By introducing couplings to light fermions, the energy in these oscillations is quickly converted into thermal radiation. We find that for a wide range of initial conditions, spanning many orders of magnitude in initial $\chi$ energy density, $\chi$ is exponentially close to the top of the tachyonic ridge by the onset of the ekpyrotic phase. 
The modified potential is sketched in Fig.~\ref{2fieldpotstab}.

\begin{figure}[htb]
\centering
\includegraphics[width=120mm]{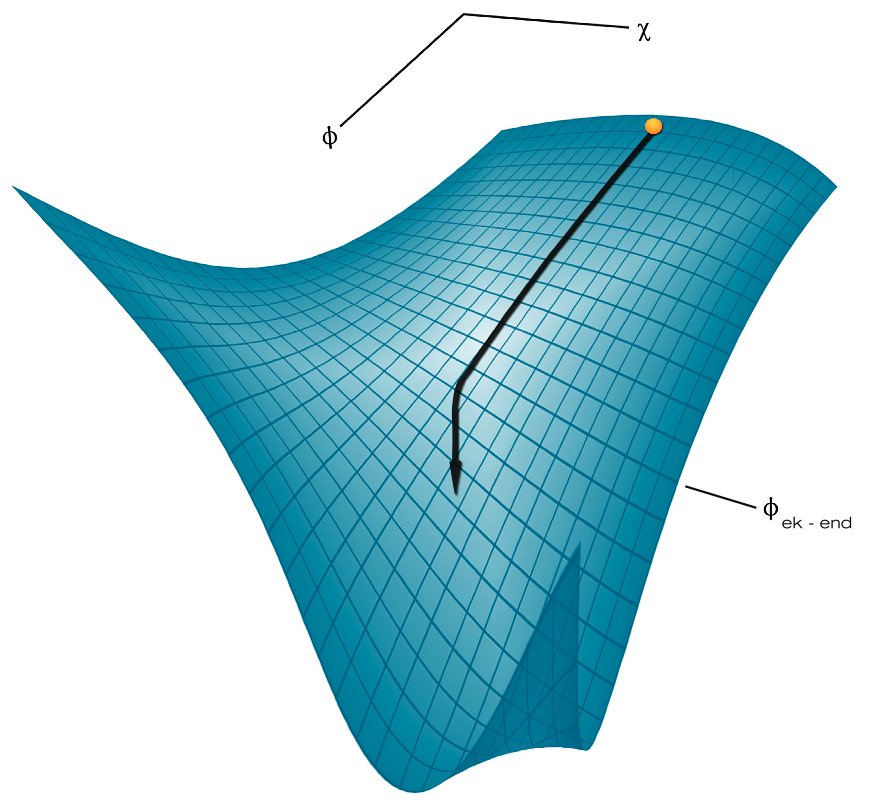}
\caption{The end of the ekpyrotic phase is triggered by a term in the potential which is small at early times but eventually pushes the field away from the tachyonic ridge. The field rolls down to a minimum in the $\chi$ direction and starts oscillating around it. Meanwhile the field keeps rolling along the $\phi$ direction.}
\label{2fieldpotend}
\end{figure}

In Sec.~\ref{end}, we turn to a discussion of the exit from the ekpyrotic phase, the subsequent NEC-violating phase and the resulting non-singular bounce. In this part of the story as well, the
($\phi,\chi$) picture leads to tremendous simplications compared to our original model~\cite{us}. To bring the ekpyrotic phase to an end, we include corrections terms which eventually drive $\chi$ away from the tachyonic ridge and towards a stable minimum, as shown in Fig.~\ref{2fieldpotend}. This is necessary to generate a turn in the field trajectory, thereby imprinting the scale-invariant entropy perturbation spectrum onto $\zeta$. 

Thus $\chi$ rolls to the minimum and starts oscillating around it. Its mass around the minimum is generically large compared to Hubble, and therefore the energy density in $\chi$ blueshifts
as a dust component. Note that the oscillations are not damped by the light fermions of the pre-ekpyrotic phase since, as $\chi$ rolls away from the tachyonic ridge, these fermions become heavy and can be henceforth ignored. Nevertheless the energy density in $\chi$ quickly becomes subdominant as $\phi$ keeps on rolling, due to the ekpyrotic attractor mechanism. The dynamics therefore effectively reduce to single-field ekpyrotic motion along $\phi$.

The rest of the story consists of $\phi$ entering a ghost condensate phase which leads to a violation of the NEC and results in a non-singular cosmological bounce. 
As described in~\cite{us}, this requires the post-ekpyrotic potential to become positive and flat. Thus, shortly after $\chi$ rolls off, we envision $\phi$ reaching a minimum of the potential followed by a steep rise to positive values where the potential becomes flat. (Note that although the potential is flat and positive, there is no sustained inflationary phase here. From the onset of the ghost condensate phase until reheating, the scale factor changes only by a factor of order unity.) Climbing up to this plateau greatly reduces the kinetic energy in $\phi$, which allows $\phi$ to enter a ghost condensate phase. While most of this bounce story parallels that of our original scenario~\cite{us}, a key difference here is that we only need one ghost condensate field, as opposed to two. This is another considerable simplification brought about by the $\phi,\chi$ framework.

The successful merger of ekpyrosis and ghost condensation leads to a number of consistency conditions which we derive explicitly in Sec.~\ref{bounce} and summarize
in Sec.~\ref{summary}.
For example, the energy density in the $\chi$ oscillations must be small enough to allow the ghost condensate to dominate the dynamics
and lead to a bounce. None of these conditions are very constraining, however, and we illustrate how they can be satisfied for a wide range of parameters
by studying a specific potential in Sec.~\ref{example}.

In the Conclusion in Sec.~\ref{conclude}, we step back and assess the current status of New Ekpyrotic Cosmology compared with the inflationary model, both at the theoretical and phenomenological levels. 

\begin{figure}[htb]
\centering
\includegraphics[width=120mm]{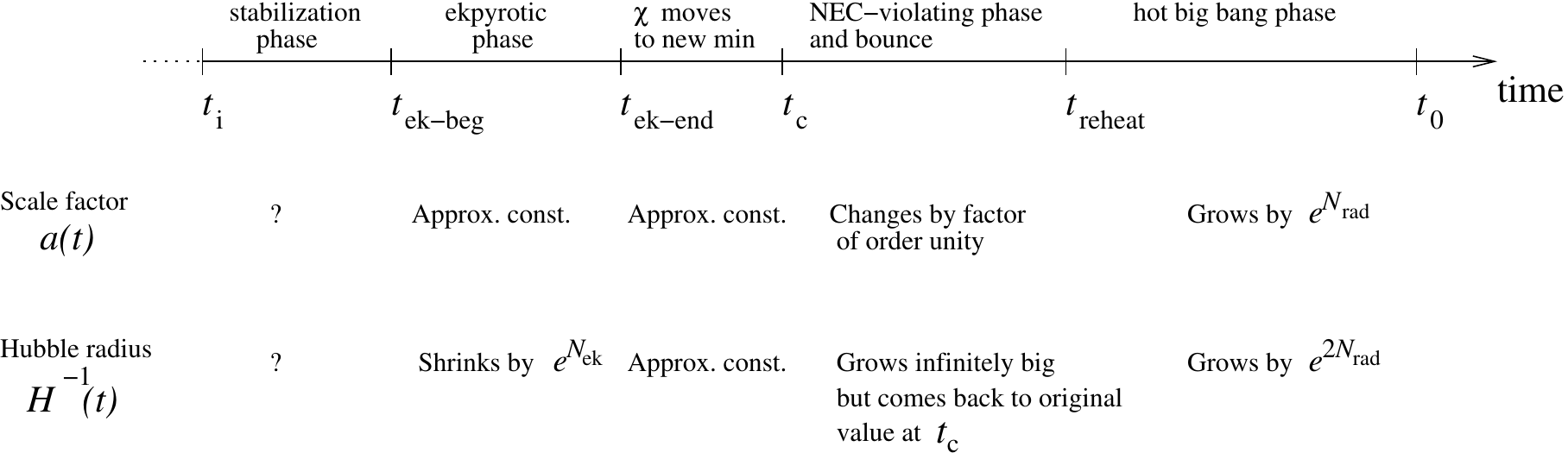}
\caption{Sequence of events in New Ekpyrotic Cosmology.}
\label{timelines}
\end{figure}

The sequence of events in New Ekpyrotic Cosmology is outlined in Fig.~\ref{timelines}. Starting from a cold initial state at some initial time $t_i$, the field $\chi$ begins oscillating around the stable minimum and decays into radiation --- see Fig.~\ref{2fieldpotstab}. Since radiation blueshifts more slowly than the energy density in $\phi$, eventually the latter takes over and dominates the energy. Meanwhile, the $\chi$ direction becomes tachyonic, allowing for the growth of entropy perturbations. This marks the onset of the ekpyrotic phase, denoted by $t_{\rm ek-beg}$. During this phase, the universe contracts very slowly, so that $a(t)$ is nearly constant, whereas the Hubble radius $H^{-1}$ shrinks by an exponential amount. It is also during this phase that fluctuations in $\chi$ are amplified and acquire a scale-invariant spectrum. The ekpyrotic phase comes to an end at $t_{\rm ek-end}$, when $\chi$ is pushed away from the tachyonic ridge and starts oscillating around a new minimum --- see Fig.~\ref{2fieldpotend}. Shortly thereafter ($t = t_c$), $\phi$ reaches a minimum in the potential, climbs up on the flat positive plateau, and enters the ghost condensate phase. This violates the NEC and leads to a non-singular bounce. Not long after the universe starts expanding, the energy density in $\phi$ is converted into matter and radiation, which reheats the universe and marks the beginning of the hot big bang phase ($t=t_{\rm reheat}$). The rest of the story until today ($t = t_0$) is standard: the universe successively undergoes epochs of radiation, matter and dark energy domination.

\section{Two-Field Ekpyrosis}
\label{2field}

The ekpyrotic mechanism for generating a pre-big bang, scale-invariant spectrum for $\zeta$,
as described in~\cite{us}, relies on two scalar fields, $\phi_1$ and $\phi_2$,
each rolling down a steep, negative and nearly exponential potential.
For concreteness, let us assume for now that the potentials are pure exponentials,
in general with different powers $p_1\ll 1$ and $p_2\ll 1$:
\be
V(\phi_1,\phi_2) =  -V_1\exp\left(-\sqrt{\frac{2}{p_1}}\frac{\phi_1}{M_{\rm Pl}}\right) -
V_2\exp\left(-\sqrt{\frac{2}{p_2}}\frac{\phi_2}{M_{\rm Pl}}\right)\,.
\label{V2field}
\ee
In this paper, $M_{\rm Pl}$ is chosen to be the ``reduced'' Planck mass where $M_{\rm Pl}=2.4\times 10^{18}$~GeV. Potentials that deviate from the pure exponential form were discussed in~\cite{us,private}.
In this section, the two scalars can be taken to have canonical kinetic terms;
all higher-derivative self-interactions, necessary to generate a ghost condensate and
produce a smooth bounce, are assumed to be negligible during this ekpyrotic phase.
Under these assumptions, $\phi_1$ and $\phi_2$ satisfy the standard scalar equations
of motion
\bea
\nonumber
& & \ddot{\phi}_1 + 3H\dot{\phi}_1 = -V_{,\phi_1} \;;\\\
& & \ddot{\phi}_2 + 3H\dot{\phi}_2 = -V_{,\phi_2} \,.
\label{eoms}
\eea
These equations, combined with the Friedmann equation, allow for a scaling solution:
\bea
\nonumber
& & a(t)\sim (-t)^{p_1+p_2}\,;\qquad H = \frac{p_1+p_2}{t}\,; \\
\nonumber
& & \phi_1(t) = \sqrt{2p_1}M_{\rm Pl}
\log\left(-\sqrt{\frac{V_1}{M_{\rm Pl}^2p_1(1-3(p_1+p_2))}} \; t\right)\,; \\
& & \phi_2(t) = \sqrt{2p_2}M_{\rm Pl}
\log\left(-\sqrt{\frac{V_2}{M_{\rm Pl}^2p_2(1-3(p_1+p_2))}} \; t\right)\,,
\label{scaling2}
\eea
where $t$ is chosen to be negative and increasing as $t \rightarrow 0$.
For $p_1,p_2\ll 1$,~\eqref{scaling2} describes a slowly-contracting universe
driven by a highly stiff fluid with equation of state
\be
w = \frac{2}{3(p_1+p_2)}-1\gg 1\,.
\ee
This cosmology is the essential feature of ekpyrotic dynamics and is 
the origin of the generation of a scale-invariant spectrum in this context.

Ekpyrotic dynamics was first studied for a single scalar field, in which case the scaling solution has the
property of being an attractor. Indeed, since $w\gg 1$ for the background,
the corresponding energy density {\it blueshifts} much faster than all other relevant components --- curvature, matter, radiation, coherent kinetic energy and
anisotropic stress. This is the precise analogue in a
contracting universe of why accelerated expansion is an attractor in an
expanding universe. In the latter context, the nearly constant potential energy of a slowly-rolling
scalar field comes to dominate because it redshifts slower than any of the above components.

\subsection{Tachyonic Instability}

In the two-field case, however, the scaling solution~(\ref{scaling2}) is {\it not}
an attractor because of a tachyonic instability in the
direction orthogonal to the field trajectory --- the so-called
entropy direction. This was first pointed out in~\cite{us,private} and studied
extensively in~\cite{wands1, wands2}. This instability is essential to the
generation of a scale-invariant spectrum of entropy perturbations
and, hence, cannot be removed~\cite{us,tolley}.
However, as we show later in this paper, it is easy to include further
terms in the potential which are negligible during the ekpyrotic phase,
but which at early times bring the field trajectory exponentially close to the desired one.

The tachyonic instability is most easily seen in terms of two new field variables, $\phi$ and $\chi$, given by
\be
\phi = \frac{\sqrt{p_1}\phi_1+\sqrt{p_2}\phi_2}{\sqrt{p_1+p_2}}\;;\qquad \chi =
\frac{\sqrt{p_2}\phi_1-\sqrt{p_1}\phi_2}{\sqrt{p_1+p_2}}\,,
\label{newfields}
\ee
The scaling solution above implies that $\chi$ remains constant on the background trajectory.
Denoting this constant by $\chi_{t}$, we find that
\be
\chi_t\equiv M_{\rm Pl}\sqrt{\frac{p_1p_2}{2(p_1+p_2)}}\log\left(\frac{p_2}{p_1}\frac{V_1}{V_2}\right)\,.
\ee
At $\chi_{t}$ the potential is tachyonic in the $\chi$ direction.
To see this, note that the potential~(\ref{V2field}) can be rewritten
in terms of $\phi$ and $\chi$ as~\cite{wands1, wands2}
\be
V(\phi,\chi) = -V_0e^{-\sqrt{\frac{2}{p}}\;\phi/M_{\rm Pl}}\left(\frac{p_1}{p}e^{-\sqrt{\frac{p_2}{p_1}}\sqrt{\frac{2}{p}}
(\chi-\chi_t)/M_{\rm Pl}}+
\frac{p_2}{p}e^{\sqrt{\frac{p_1}{p_2}}\sqrt{\frac{2}{p}}(\chi-\chi_t)/M_{\rm Pl}}\right)\,,
\label{Vfull}
\ee
where $p\equiv p_1+p_2$ and $V_0\equiv (1+p_2/p_1)V_1\exp\left(-\sqrt{p_2/p_1p}\;\chi_t/M_{\rm Pl}\right)$.
Taylor-expanding~\eqref{Vfull} as a power series in $\chi-\chi_{t}$, we find to quadratic order that
\be
V(\phi,\chi) = -V_0e^{-\sqrt{\frac{2}{p}}\;\phi/M_{\rm Pl}}\left(1+\frac{1}{pM_{\rm Pl}^2}(\chi-\chi_t)^2+\ldots\right)\,,
\label{Vapprox}
\ee
thereby revealing a tachyonic mass squared 
in the $\chi$ direction.
This tachyonic ridge is sketched in Fig.~\ref{2fieldpotek}.
Note that~(\ref{Vapprox}) satisfies 
\begin{equation}
V_{,\chi\chi} = V_{,\phi\phi}
\label{cat1}
\end{equation}
at $\chi=\chi_t$, an essential feature in generating a scale-invariant
spectrum for $\delta\chi$, as we review below. 

In terms of $\phi$ and $\chi$, the scalar equations of motion~(\ref{eoms}) take the form
\bea
\nonumber
& & \ddot{\phi} + 3H\dot{\phi}= -V_{,\phi} \;;\\\
& & \ddot{\chi} + 3H\dot{\chi} = -V_{,\chi} \,,
\label{eomsphichi}
\eea
while the background solution~(\ref{scaling2}) becomes
\bea
\nonumber
& & a(t)\sim (-t)^p\,;\qquad H = \frac{p}{t}\,; \\
\nonumber
& & \phi(t) = \sqrt{2p}M_{\rm Pl}\log\left(-\sqrt{\frac{V_0}{M_{\rm Pl}^2p(1-3p)}}\; t\right)\, ;\\
& & \chi = \chi_t \, .
\label{5.2}
\eea
This describes the system rolling along the $\phi$ direction on the top of the tachyonic ridge in $\chi$
with the effective scalar potential for $\phi$ given by
\be
V_{\rm eff}(\phi) = -V_0\exp\left(-\sqrt{\frac{2}{p}}\frac{\phi}{M_{\rm Pl}}\right)\,.
\label{hello}
\ee
Note, using~\eqref{Vapprox},~\eqref{5.2} and~\eqref{hello}, that
\begin{equation}
V_{,\phi\phi}|_{\chi=\chi_{t}}=V_{{\rm eff}, \phi\phi} \approx -\frac{2}{t^{2}}
\label{cat2}
\end{equation}
for $p\ll 1$. As $\phi$ rolls down this steep, negative exponential potential, the tachyonic mass for $\chi$ gets larger in magnitude.
A useful relation between the latter and other background quantities is
\be
m_{\rm tachyon}^2 \equiv -V_{,\chi\chi}|_{\chi=\chi_{t}} \approx 2\frac{H^2}{p^2}\,,
\label{mtachyon}
\ee
where we have used~\eqref{cat1},~\eqref{5.2} and~\eqref{cat2}. Since $p\ll 1$, the rate of instability is always much greater than the Hubble parameter.

\subsection{Instability and Scale-Invariant Entropy Perturbation}
\label{pertrev}

Since the field trajectory is orthogonal to $\chi$, the fluctuations in the latter by definition coincide  during the ekpyrotic phase with the entropy perturbations. Moreover the tachyonic mass uncovered above is essential in amplifying these perturbations, as we now review. This implies that the instability
cannot be cured without spoiling the scale invariance of the entropy perturbation spectrum. 

The evolution 
of the Fourier modes of $\delta\chi$ is governed by~\cite{us}
\be
\ddot{\delta \chi_k} + 3H\dot{\delta \chi_k} + \left(\frac{k^2}{a^2}+V_{,\chi\chi}\right)\delta \chi_k = 0\,.
\label{dchi}
\ee
In Sec.~\ref{tilt} we will solve~(\ref{dchi}) and derive a general expression for the spectral tilt, 
including Hubble damping and corrections from the potential.
For the moment, we note from~\eqref{mtachyon} that
\begin{equation}
\frac{H}{\sqrt{|V_{,\chi\chi}|}} \approx \frac{p}{\sqrt{2}}
\label{hello2}
\end{equation}
at $\chi=\chi_{t}$. Thus, since $p\ll1$, we can neglect the Hubble damping term in~\eqref{dchi} and treat space as flat --- we can set $a=1$ without loss of generality. In this approximation, $\chi$ is a free scalar with time-dependent mass:
\be
\ddot{\delta \chi_k} + \left(k^2 -\frac{2}{t^2}\right)\delta\chi_k = 0\,,
\label{dchiflat}
\ee
where we have used~\eqref{cat1} and~\eqref{cat2} to replace $V_{,\chi\chi}$ by $-2/t^{2}$. 

Well within the horizon, a given $k$-mode starts out in the usual Bunch-Davies vacuum:
$\delta\chi_k= e^{-ikt}/\sqrt{2k}$. The solution to~(\ref{dchiflat}) with these initial conditions is
\be
\delta\chi_k = \frac{e^{-ikt}}{\sqrt{2k}}\left(1-\frac{i}{kt}\right)\,.
\label{exactdchi}
\ee
On super-Hubble scales, $k(-t)\ll 1$, up to an irrelevant phase factor this reduces to
\be
k^{3/2}\delta\chi_k = \frac{1}{\sqrt{2}(-t)}\,,
\label{again}
\end{equation}
corresponding to a Harrison-Zeldovich spectrum. It follows that, 
as stated above, the scale invariance of the entropy spectrum in 
two-field ekpyrosis relies on the existence of a tachyonic instability 
along one field direction, denoted by $\chi$ above. 

\section{Generalized Ekpyrotic Potentials}
\label{simplified}

From the above analysis we learn two important lessons about the ekpyrotic phase: first, the physics associated with two exponential potentials for $\phi_1$ and $\phi_2$ is most transparent when the theory is rewritten in terms of the rotated fields $\phi$ and  $\chi$; second, that a nearly scale-invariant spectrum arises in the fluctuations of $\chi$ as a result of the relations $V_{,\phi\phi} \approx -2/t^2$ and $V_{,\phi\phi} =V_{,\chi\chi}$ at $\chi =\chi_{t}$ in potential~\eqref{Vapprox}. From the first fact we conclude that it is far simpler to treat $\phi$ and $\chi$ as the fundamental fields in the theory, which we do henceforth. It follows from the second fact that we are free to modify the potential 
in the ekpyrotic phase as long as the relations
\begin{equation}
V_{,\phi\phi} \approx -2/t^2\,,  \qquad V_{,\phi\phi} \approx V_{,\chi\chi} 
\label{cat3}
\end{equation}
at $\chi =\chi_{t}$ continue to hold. This new perspective greatly simplifies New Ekpyrotic Cosmology and broadens the class of allowed potentials.

\subsection{Generalized Potentials}

The simplest such modification is the following. Motivated by~\eqref{Vapprox}, consider any potential of the form
\be
V(\phi,\chi) = -V_0e^{-\sqrt{\frac{2}{p}}\;\phi/M_{\rm Pl}}\left(1+\frac{1}{pM_{\rm Pl}^2}(\chi-\chi_t)^2 + F(\phi,\chi-\chi_{t})\right)\,,
\label{Vbew}
\ee
where $\chi_t$ is any constant. Meanwhile, the function $F$ satisfies
\begin{equation}
F_{,\chi\chi}\vert_{\chi=\chi_t}=0
\label{cat4}
\end{equation}
but is otherwise arbitrary. Since the exponential in $\phi$ and the quadratic term in the  Taylor expansion in $\chi-\chi_{t}$ are identical to those in~\eqref{Vapprox}, this modified potential clearly continues to satisfy conditions~\eqref{cat3}. Note that in the case of two exact exponentials discussed above, keeping higher order terms in the expansion~\eqref{Vapprox} gives
\begin{equation}
F(\phi,\chi-\chi_{t})= \frac{\sqrt{2}(p_1-p_2)}{3p^{3/2}\sqrt{p_1p_2}M_{\rm Pl}^3}(\chi-\chi_t)^3 + \frac{p_1^3+p_2^3}{6p_1p_2p^3M_{\rm Pl}^4}(\chi-\chi_t)^4 + \dots,
\label{cat5}
\end{equation}
which manifestly satisfies~\eqref{cat4}. Hence,~\eqref{Vapprox} is indeed of the form~\eqref{cat4}. Since the generalized functions $F$ encode the self-interactions for $\chi$, they play an important role in the analysis of non-Gaussianity in New Ekpyrotic  Cosmology, which we discuss below. 

The ($\phi,\chi$) perspective suggests an even broader class of allowed ekpyrotic potentials. Note that although the exponential in $\phi$ and the quadratic term in the Taylor expansion in $\chi-\chi_{t}$ imply that the potential~\eqref{Vbew}  satisfies conditions~\eqref{cat3}, they are hardly the only functions to do so. For simplicity of notation, let us henceforth set 
\begin{equation}
\chi_t=0
\label{aa}
\end{equation}
without loss of generality. Then, consider any potential of the form
\be
V(\phi,\chi) = \cV(\phi)\left(1+\frac{1}{2}\frac{\cV_{,\phi\phi}}{\cV}f(\phi)\chi^2 + F(\phi,\chi)\right)\,,
\label{genpot}
\ee
where $\cV(\phi)$ satisfies the standard fast-roll conditions of single-field ekpyrosis
\bea
\nonumber
\epsilon &\equiv& M_{\rm Pl}^{-2}\left(\frac{\cV}{\cV_{,\phi}}\right)^2 \ll 1\\
\eta &\equiv& 1-\frac{\cV\cV_{,\phi\phi}}{\cV_{,\phi}^2}\,; \quad |\eta| \ll 1\,,
\label{3.3}
\eea
which require the potential ${\cal{V}}(\phi)$ to be steep and nearly exponential, respectively.
Meanwhile, the function $f(\phi)$ has the property
\begin{equation}
f(\phi) \approx 1
\label{cat6}
\end{equation}
over the relevant range of modes, and $F$ is constrained by condition~\eqref{cat4}.
It is straightforward to show that potential~\eqref{genpot} satisfies the conditions~\eqref{cat3} for a nearly scale-invariant spectrum and is the most general potential to do so.
Clearly the potential in~(\ref{Vbew}) --- with $\chi_t$ set to zero --- is of this form, with
$\cV(\phi) = -V_0e^{-\sqrt{2/p}\;\phi/M_{\rm Pl}}$ and $f(\phi) \equiv 1$.

One might object that, at first sight, the $(\phi,\chi)$ perspective requires an unnatural fine-tuning --- $f(\phi) \approx 1$ ---
while this is seemingly automatically satisfied in the original $\phi_1,\phi_2$ language. This would seem to suggest that
the latter is a more convenient set of fundamental fields to study. On the contrary, we argue that the original picture in terms of $\phi_1$ and $\phi_2$
requires much more fine-tuning. The generalized potential $V(\phi,\chi)$ in~(\ref{genpot}) requires one function to be nearly exponential and one function in the Taylor expansion --- $f(\phi)$ --- to be nearly unity. However, the shape of the potential away from $\chi= 0$, given by the general function $F$, can be quite arbitrary. In contrast, in the $\phi_1$ and $\phi_2$ framework a second function must be of nearly exponential form, corresponding to tuning infinitely many coefficients in a Taylor expansion.

The generalized function ${\cal{V}}(\phi)$ and the modified quadratic term in the Taylor expansion in $\chi$ have a profound impact on the form of the spectral index of the theory. This will now be discussed in detail.

\subsection{Spectral Index for General Potentials}
\label{tilt}

In this section we derive the spectral tilt for the general potential~(\ref{genpot}). A necessary condition for scale invariance, as we will see, is that the function
$f(\phi)$ be approximately unity over the relevant range of modes, which ensures that $V_{,\chi\chi}\approx V_{,\phi\phi}$ near $\chi=\chi_t=0$.
Therefore, let us parametrize $f$ as
\begin{equation}
f(\phi)=1 + 3 \delta\,; \qquad |\delta | \ll 1\,, \label{6.3}
\end{equation}
where the factor of $3$ is introduced for convenience. Note that, generically, $\delta$ can be a function of $\phi$. For comparison, the analysis of Sec.~4 in~\cite{us} focused on the special case of identical potentials for $\phi_1$ and $\phi_2$,
which translates in the $\phi,\chi$ variables to $f(\phi)=1$. Turning on a non-zero $\delta$ is not only more general but, as we will see, allows for greater freedom in the spectral tilt.

The perturbation equation for the entropy field $\delta \chi$ is given in~\eqref{dchi}  by
\begin{equation}
\ddot{\delta \chi_k}+ 3 H \dot{\delta \chi_k}
+\left(\frac{k^2}{a^2}+V_{,\chi\chi}\right) \delta \chi_k =0\,.
\label{6.4}
\end{equation}
When $\delta=0$, we find from~\eqref{genpot} that $V_{,\chi
\chi}=V_{,\phi \phi}$ at $\chi=0$, in which case the calculation is
identical to that presented in~\cite{us, private}. When $\delta \neq 0$, however,
$V_{, \chi \chi}= V_{, \phi \phi}(1 + 3 \delta)$ at $\chi=0$ and thus 
\begin{equation}
\ddot{\delta \chi_k}+ 3 H \dot{\delta \chi_k}
+\left(\frac{k^2}{a^2}+V_{,\phi \phi} (1+3 \delta) \right)
\delta \chi_k =0. \label{6.7}
\end{equation}

To calculate $n_s$, we follow the approach developed in~\cite{us}.  Define the equation of state parameter
\begin{equation}
\bar{\epsilon} \equiv \frac{3}{2}(1+w)=-\frac{\dot{H}}{H^2}=-\frac{d \ln
H}{d N}\,, \label{6.8}
\end{equation}
where $N=\ln a$ is a dimensionless time variable. Equation~\eqref{6.7} can now be recast entirely in terms of
$\bar{\epsilon}$ and its derivatives with respect to $N$, as shown in~\cite{us} for the $\delta = 0$ case. Since the calculation
closely parallels that of~\cite{us}, for the sake of brievity we only outline the key steps and refer the reader to~\cite{us} for details.

In the ekpyrotic phase, the equation of state parameter is large: $\bar{\epsilon}\gg 1$. Moreover, for simplicity, we assume that it is slowly-varying in time.
In this limit, the $\epsilon$ and $\eta$ fast-roll parameters can be written as
\begin{equation}
\epsilon =\frac{1}{2 \bar{\epsilon}}\,; \qquad \eta =\frac{1}{4
\bar{\epsilon}^2} \frac{d \bar{\epsilon}}{d N}\,. \label{6.9}
\end{equation}
Terms involving higher-derivatives on $\bar{\epsilon}$, such as $(d \bar{\epsilon}/d N)^2$ and $d^2 \bar{\epsilon}/d N^2$, are found to be higher-order
in  $\epsilon$ and $\eta$ and, therefore, can be consistently neglected in the fast-roll approximation. To illustrate how this substitution works,
we note, for example, that $V_{,\phi\phi}$ in~(\ref{6.7}) can be rewritten as
\begin{equation}
\frac{V_{,\phi \phi}}{H^2} =
\bar{\epsilon}^2\left(-2+\frac{6}{\bar{\epsilon}} +\frac{5}{2}
\frac{1}{\bar{\epsilon}^2}\frac{d \bar{\epsilon}}{d N}+\dots\right) =
\bar{\epsilon}^2(-2+ 12 \epsilon + 10 \eta + \dots)\,, \label{6.10}
\end{equation}
where the ellipses stand for terms of order $\epsilon^2, \eta^2, \epsilon \eta$ and so on. Let us emphasize that $\epsilon$ and $\eta$ can therefore be treated as essentially constant in this approximation, as in slow-roll inflation. Their time-dependence generate order $\epsilon^2, \eta^2, \epsilon \eta$ corrections, which are negligible.

The next step consists of introducing a new time variable
\begin{equation}
x=\frac{1}{\bar{\epsilon}-1} \left(\frac{k}{a H}\right)
\label{6.11}
\end{equation}
and rescaling $\delta \chi_k$ as
\begin{equation}
v_k= a \;\delta \chi_k \,. 
\label{6.12}
\end{equation}
After some algebra, the perturbation equation~\eqref{6.7} can be cast in the form
\begin{equation}
x^2 \frac{d^2 v_k}{d x^2} +\frac{x^2}{1-8 \eta} v_k-2 \left(1-3
(\epsilon -\eta -\delta)\right)v_k=0\,, \label{6.13}
\end{equation}
where we have dropped higher-order terms in the fast-roll parameters.

Thus far, $\epsilon$ and $\eta$ are approximated as constant whereas $\delta$ is up to this point an arbitrary function of $\phi$, and thus an arbitrary function of time.
In order for the spectrum to be nearly scale-invariant, however, clearly $\delta$ cannot vary rapidly while the observable modes exit the horizon. A small
time-dependence is allowed, of course, which would lead to a running spectral index. We leave this interesting possibility aside and henceforth focus on the simplest case where 
$\delta$ is approximately constant over the relevant spectral range. 

In terms of $y\equiv x/\sqrt{1-8\eta}$, the solution with Bunch-Davies vacuum is given as usual by a 
Hankel function:
\be
v_k = \frac{1}{\sqrt{2k}}\sqrt{\frac{\pi}{2}}\sqrt{y} H_{n}^{(1)}(y)\,,
\label{vsoln}
\ee
with $n  \equiv (3/2)\sqrt{1-8(\epsilon-\eta-\delta)/3}\approx 3/2-2(\epsilon-\eta-\delta)$. 
Note that in the limit of flat space ($a\rightarrow 1$) and 
exact scale-invariance ($n\rightarrow 3/2$), this
reduces to~(\ref{exactdchi}).

On large scales, $k\ll aH$, the amplitude tends to $v_k \sim k^{-n}$, corresponding to a spectral index:
\be
n_s-1 =4 (\epsilon - \eta -\delta)\,. \label{6.14}
\end{equation}
As a check, for $\delta=0$ this reduces to the result of~\cite{us}.
Thus, $\delta$ acts effectively as a correction to the fast-roll parameters contributing to the
spectral tilt.

As discussed in~\cite{us,private,paologhost}, pure exponential potentials --- $\eta=\delta=0$ in our language --- yield a blue spectrum.
However, as shown in~\cite{us,private} for the case $\delta=0$, it is easy to get a red tilt
by considering potentials that deviate from the pure exponential form, corresponding to non-zero $\eta$. For example, $V(\phi) \sim \exp(-\phi^n)$, with $n>2$,
gives a red tilt at large $\phi$. What we have found here is that the more generic case with $\delta\neq 0$ allows greater freedom in getting a red tilt. Even if $\epsilon-\eta$ is positive, for instance, which by itself would yield a blue spectral index, it is still possible to get a red tilt if $\delta$ is positive and dominant.

\subsection{Non-Gaussianity in General Potentials}
\label{nongauss}

Having derived the spectral index for the generalized potential~\eqref{genpot}, we now turn to the question of non-Gaussianity of the fluctuations. Whereas the spectral index is determined by the $\phi$ dependent over-all factor $\cV(\phi)$ and the form of the quadratic term in the Taylor expansion in $\chi$, the level of non-Gaussianity is associated with the function $F$, which encodes
all self-interaction terms for $\chi$. Here we provide an estimate of the non-Gaussianity level based on the generic size of these interaction terms. For simplicity, we shall ignore the effects of gravity,
as we did in Sec.~\ref{pertrev} to estimate the $\delta\chi$ 2-point function, and work to leading order in the fast-roll parameters $\epsilon$ and $\eta$. 

While it is clear from the above discussion that $F$ is completely arbitrary, nevertheless we can give some naturalness arguments for the cubic and higher-interaction terms in $\chi$ based on the form
of the mass term. To leading order in $\epsilon, \eta$, it follows from~\eqref{3.3} that 
\be
\frac{\cV_{,\phi\phi}}{\cV}\approx \epsilon^{-1}\,.
\label{V''V}
\ee
Thus the quadratic term in~(\ref{genpot}) can be written as 
\be
\frac{1}{2}\cV(\phi(t)) \frac{\chi^2}{\Lambda^2}\,,
\ee
where $\Lambda$ is an effective cut-off scale: 
\be
\Lambda \equiv M_{\rm Pl}\sqrt{\epsilon}\,. 
\label{Lamdef}
\ee

From this point of view, we expect higher-order interactions to be suppressed by the same scale, leading to a generic $F$ of the form
\be
F(\phi,\chi) = \frac{\alpha(\phi(t))}{3}\frac{\chi^3}{\Lambda^3} + \cO\left(\chi^4\right)\,.
\ee 
Here the coefficient $\alpha$ is {\it a priori} an arbitrary function of $\phi$. However, to simplify the analysis we shall take 
it to be a constant which, naturally,  is expected to be of order unity.

The three-point function for the fluctuations $\delta\chi$ is determined by the interaction Hamiltonian~\cite{malda}, which we can read off from the cubic part of the potential:
\be
H_{\rm int} = \cV(\phi(t)) \frac{\alpha}{3}\frac{\delta\chi^3}{\Lambda^3}  \approx -\frac{\alpha}{3\Lambda t^2}\delta\chi^3\,.
\ee
Note that in the last step we have substituted~(\ref{V''V}) and used $\cV_{,\phi\phi} \approx -2/t^2$. Since $H_{\rm int}$ is exactly of the form studied in~\cite{paologhost}, we can immediately
apply their results to our calculation. We find that the 3-point function for the curvature perturbation $\zeta$ at the end of the ekpyrotic phase
($t=t_{\rm ek-end}$), inherited from $\delta\chi$, is given by
\be
\langle \zeta_{\vec{k}_1}(t_{\rm ek-end}) \zeta_{\vec{k}_2}(t_{\rm ek-end})\zeta_{\vec{k}_3}(t_{\rm ek-end})\rangle \approx (2\pi)^3\delta\left(\sum_i \vec{k}_i\right)\frac{\sum_ik_i^3}{\prod_ik_i^3} \frac{\alpha\Delta_\zeta^{3/2}}{2^{3/2}\Lambda(-t_{\rm ek-end})}\,,
\label{zeta3pt}
\ee
where $k_i$ denotes the magnitude of $\vec{k}_i$, and all modes are assumed to be super-Hubble: $k_i(-t_{\rm ek-end}) \ll 1$. 
Moreover, $\Delta_\zeta$ is the amplitude of the power spectrum, $\Delta_\zeta = k^3\zeta_k^2$,
which is fixed by WMAP: 
\be
\Delta_\zeta^{1/2} \approx 10^{-5}\,. 
\label{COBE}
\ee
Note that the shape of the 3-point function is of the {\it local} form~\cite{shape}.

The level of non-Gaussianity is usually expressed in terms of the $f_{\rm NL}$ parameter~\cite{komat1,komat2}, defined in terms of $\zeta$ by
\be
\zeta(x) = \zeta_g(x) - \frac{3}{5}f_{\rm NL}\zeta_g^2(x)\,,
\ee
where $\zeta_g$ has a Gaussian spectrum. From~(\ref{zeta3pt}) we immediately read off that
\be
f_{\rm NL} = \mp \frac{5\sqrt{2}}{24}\Delta_\zeta^{-1/2} \frac{\alpha}{\Lambda(-t_{\rm ek-end})} \approx \mp \frac{5}{24}\Delta_\zeta^{-1/2} \frac{\alpha}{\epsilon}\cdot\frac{|H_{\rm ek-end}|}{\sqrt{2\epsilon}M_{\rm Pl}}\,,
\label{fNL}
\ee
where we have substituted~(\ref{Lamdef}) and~(\ref{5.2}), taking into account that $p\approx 2\epsilon$ in our approximation. Note that the above sign ambiguity has to do with the details of how the entropy perturbation is converted onto $\zeta$~\cite{paologhost}, an issue we ignore for the present discussion.

The last factor in~(\ref{fNL}) can be expressed in terms of the amplitude of density perturbations as~\cite{us}
\be
\Delta_\zeta^{1/2} = \beta \frac{H_{\rm reheat}}{\sqrt{2\epsilon}M_{\rm Pl}}\,,
\label{Delzeta}
\ee
where $\beta$ is a model-dependent prefactor having to do with the details of the exit from the ekpyrotic phase. As we will see in Sec.~\ref{end}, in the approximation that the ekpyrotic phase
ends through a sharp turn in the field trajectory, then we have $\beta=\Delta\theta$, where $\Delta\theta$ is the change in angle in the trajectory in field space --- see also~(5.5) of~\cite{us}. Thus $\beta$ is at most of order unity in that case. 

Save for the $\beta$ coefficient, this expression is identical to its counterpart in inflation,
with $\epsilon$ replacing the $\epsilon_{\rm inf}$ slow-roll parameter. Now, as indicated in Fig.~\ref{timelines} and as we will discuss later in detail, the magnitude of the Hubble parameter at the end of the ekpyrotic phase is roughly the same as at reheating: $|H_{\rm ek-end}| \sim H_{\rm reheat}$. Using this fact and substituting~(\ref{Delzeta}) into~(\ref{fNL}), we find
\be
f_{\rm NL}  \approx \mp \frac{5}{24}\frac{\alpha}{\beta}\;\epsilon^{-1}\,.
\ee
At the 2$\sigma$ level, the recent WMAP data constrains $f_{\rm NL}$ to be within the range~\cite{wmap3,paolonongauss}: $-36 < f_{\rm NL} < 100$. Using the liberal end of this bound, we get
\be
\epsilon \;\gsim\; 2\cdot 10^{-3}\;\frac{\alpha}{\beta}\,.
\ee
Since $\epsilon \ll 1$, the level of non-Gaussianity tends naturally to be large in New Ekpyrotic Cosmology. For example, we will see in Sec.~\ref{numbers} that $p \sim10^{-2}$ for $T_{\rm reheat}  = 10^{15}$~GeV, corresponding to $\epsilon \approx 5\cdot 10^{-3}$. Furthermore, as discussed in Sec.~\ref{end}, typically $\beta\sim\cO(1)$. Hence, taking  $\alpha\sim\cO(1)$ yields $f_{\rm NL}\approx 40$, which is below, but not far from, the present upper bound. Lower reheating temperatures correspond to even smaller values of $\epsilon$ and therefore to larger non-Gaussian signal. 
Of course, we are free to choose the parameter $\alpha$ to be small and, hence, ekpyrotic theory can always lie within the observational bound. The implications for current and future experiments will be discussed elsewhere.

The significant non-Gaussian signal is a distinguishing feature of New Ekpyrosis compared to slow-roll inflation, as the latter is well-known to give unobservably small $f_{\rm NL}$. Significant non-Gaussianity in inflation is achieved in multi-field models~\cite{uzan}, in models with higher-derivative kinetic terms~\cite{kinf,gary,DBI,ghostinf}, as well as in scenarios where density perturbations are generated by another light field, such as the curvaton~\cite{curvaton} and modulon scenarios~\cite{modulon}. The large non-Gaussianity level of New Ekpyrotic Cosmology is also in sharp contrast with single-field ekpyrosis, where density perturbations are even more Gaussian that in slow-roll inflation~\cite{davidpaul}.

\section{Flatness and Homogeneity in Ekpyrotic Theory}
\label{flatness}

The tachyonic instability in $\chi$ is required in order to generate a nearly scale-invariant fluctuation spectrum during the ekpyrotic phase. Its existence, however, necessitates a careful study of both the initial conditions of ekpyrotic cosmology, as well as the mechanism for ending the ekpyrotic phase and generating a Harrison-Zeldovich spectrum in the curvature perturbation. We will give a detailed analysis of both of these issues later in this paper. Here, however, we note that initial conditions are closely related to the required degree of flatness and homogeneity of the universe. In this section, we will show that both of these fundamental problems are naturally solved in New Ekpyrotic Cosmology.

When the original ekpyrotic scenario was proposed, it was believed that the model failed to address the flatness and homogeneity problems of standard big bang cosmology. This was viewed as a drawback for ekpyrotic theory as compared to inflation~\cite{pyro,lindenew}. This concern motivated the cyclic extension of the scenario~\cite{cyc,design}, where flatness and homogeneity at the onset of each contracting phase is achieved by an epoch of dark energy domination at the conclusion of the
previous cycle. However, much progress has been made since then and the time seems right to revisit these issues in the pure, non-cyclic ekpyrotic context. We will argue that, contrary to initial beliefs, ekpyrotic theory is equally successful as inflation in addressing the standard problems of big bang cosmology~\cite{cyclicages}, without having to invoke any phase of cosmic acceleration. 

\subsection{Contrast with Inflation}

Inflationary and ekpyrotic cosmology are based on starkly different assumptions about the initial state of the universe. 

In inflation, one envisions the universe as starting at the big bang in a hot, rapidly expanding and chaotic state. While most of the universe is wildly inhomogeneous,
the assumption is that there exists a Hubble patch somewhere which is sufficiently smooth to allow inflation to occur. A short phase of cosmic acceleration then
blows up this tiny region into a large, homogeneous, isotropic and flat universe. 

Ekpyrosis, on the other hand, proposes that the universe begins in a quiescent, nearly vacuous and cold state. In the context of the original ekpyrotic scenario,
this starting point was motivated by nearly supersymmetric or BPS initial conditions. In this scenario, all physical scales at the onset of ekpyrosis --- initial Hubble parameter, curvature of the universe, radiation temperature --- are very small in absolute, particle physics terms. 

In the absence of a concrete theory of initial conditions, however, the question of whether the universe started out in a hot or cold state belongs to the realm of metaphysics. Thus, with our current understanding, a tiny initial Hubble parameter is no less natural than a GUT-scale Hubble. From a cosmology perspective, the relevant yardstick for assessing the degree of genericity of initial conditions should, therefore, not be based on the absolute scale of any observable, such as the initial spatial curvature, but rather on whether it is fine-tuned relative to other observables, such as the energy density in other components. We illustrate this point by discussing the flatness and homogeneity problems of standard big bang cosmology. 

\subsection{Flatness and Homogeneity Problems in Inflation}

Viewed as an additional energy component in the Friedmann equation, the curvature of the universe contributes a fractional amount
\be
\Omega_k \sim \frac{1}{a^2H^2}\,.
\ee
Since both $a$ and $H$ evolve in time, evidently so does $\Omega_k$. In particular, from the onset of the hot big bang phase until the present time the scale factor grows by a factor of $e^{N_{\rm rad}}$. Now, assuming for simplicity that our universe has been radiation-dominated ever since reheating,  we have $H \sim 1/t \sim 1/a^2$. It follows that over the same period the Hubble parameter shrinks by $e^{-2N_{\rm rad}}$. In other words, $aH$ changes by $e^{-N_{\rm rad}}$.  A useful expression for $N_{\rm rad}$ is therefore
\be
N_{\rm rad} \equiv \ln\left(\frac{a_{\rm reheat}H_{\rm reheat}}{a_0H_0}\right) = \ln\left(\frac{T_{\rm reheat}}{T_0}\right)\,,
\label{Nrad}
\ee
where the last step follows from the standard redshift relation $a\sim 1/T$. For instance, since the current temperature
of the universe is $T_0 = 2.7 K \sim 10^{-3}$~eV, then an initial temperature of $T_{\rm reheat}  = 10^{15}$~GeV yields the standard
$N_{\rm rad} \approx 60$, whereas a lower reheat temperature of $10^8$~GeV corresponds to $N_{\rm rad} \approx 46$. 
Incidentally, we immediately recognize $e^{N_{\rm rad}}$ as the ratio between the size of our observable
Hubble patch in comoving units ($\lambda_0 = H^{-1}_0/a_0$) to that at reheating ($\lambda_{\rm reheat} = H^{-1}_{\rm reheat}/a_{\rm reheat}$).

Therefore, during the radiation-dominated epoch $\Omega_k$ grows by an exponential factor:
\be
\frac{\Omega_k^{(0)}}{\Omega_k^{({\rm reheat})}} = \frac{a_{\rm reheat}^2H_{\rm reheat}^2}{a_0^2H_0^2} = e^{2N_{\rm rad}}\,.
\label{omkinf}
\ee
Since we know from observations that $\Omega_k$ is at most a few percent today, we are forced to conclude that $\Omega_k$ must have been exponentially
small relative to all other components at the onset of the hot big bang phase. This is the flatness problem.

Inflation addresses this fine-tuning through a short phase of cosmic acceleration. During this phase the scale factor grows by an exponential amount, 
whereas the Hubble parameter remains essentially constant. Thus we can define the total number of e-foldings of inflationary expansion from some initial
time $t_i$ until reheating $t_{\rm reheat}$ as
\be
N_{\rm inf} \equiv \ln\left(\frac{a_{\rm reheat}H_{\rm reheat}}{a_iH_i}\right)\approx  \ln\left(\frac{a_{\rm reheat}}{a_i}\right)\,.
\ee
By definition $e^{N_{\rm inf}}$ measures the size of the proto-inflationary patch in comoving units ($\lambda_i = H^{-1}_i/a_i$) to that at reheating.
 
This immediately implies that any initial curvature is exponentially suppressed by the time of reheating:
\be
\frac{\Omega_k^{({\rm reheat})}}{\Omega_k^{(i)}} = \frac{a_i^2H_i^2}{a_{\rm reheat}^2H_{\rm reheat}^2} = e^{-2N_{\rm inf}}\,.
\ee
Then, as long as $N_{\rm inf} \;\gsim\; N_{\rm rad}$, we find that $\Omega_k^{(0)}\;\lsim\; \Omega_k^{(i)}$. In other words, 
starting from the natural choice $\Omega_k^{(i)}\sim \cO(1)$, a long enough inflationary phase results in a sufficiently large radius of curvature
for the observable universe. This is the essence of how inflation solves the flatness problem. Instead of requiring an exponentially large 
initial radius of curvature, as in standard big bang cosmology, inflation tolerates an initial curvature component comparable to all other forms of
energy density. 

The homogeneity problem follows automatically from the above considerations, provided that space-time is essentially
homogeneous within the proto-inflationary patch. If $N_{\rm inf} \;\gsim\; N_{\rm rad}$, then our entire observable universe
lies well-within this patch at the onset of inflation and is therefore highly homogeneous.

\subsection{Flatness and Homogeneity Problems in Ekpyrotic Cosmology}
\label{flathomogek}

Ekpyrotic theory solves the flatness and homogeneity problems in an analogous way, yet using drastically different dynamics. Instead of exponentially rapid expansion with nearly constant Hubble radius, the ekpyrotic phase consists of a slowly contracting universe with rapidly shrinking Hubble radius. The scale factor is essentially constant during this phase, as indicated in Fig.~\ref{timelines}. Meanwhile, the Hubble radius shrinks by an overall exponential factor $e^{N_{\rm ek}}$, with $N_{\rm ek}$ defined by
\be
N_{\rm ek} \equiv \ln\left(\frac{a_{\rm ek-end}H_{\rm ek-end}}{a_{\rm ek-beg}H_{\rm ek-beg}}\right) \approx \ln\left(\frac{H_{\rm ek-end}}{H_{\rm ek-beg}}\right)\,.
\label{Nek}
\ee
This definition implies that $e^{N_{\rm ek}}$ is just the ratio of the comoving Hubble radius at the onset and exit of the ekpyrotic phase. 
Thus by the end of the ekpyrotic phase the curvature component is suppressed by
\be
\frac{\Omega_k^{({\rm ek-end})}}{\Omega_k^{({\rm ek-beg})}} =  \left(\frac{a_{\rm ek-beg}H_{\rm ek-beg}}{a_{\rm ek-end}H_{\rm ek-end}}\right)^2  = e^{-2N_{\rm ek}}\,.
\ee

This is followed by a bounce and an expanding hot big bang phase. In the New Ekpyrotic scenario, this is achieved by a ghost condensation phase that violates the NEC, generates a non-singular bounce and eventually reheats the universe. As indicated  in Fig.~\ref{timelines}, however, between $t_{\rm ek-end}$ and $t_{\rm reheat}$ the scale factor changes by at most a factor of order unity. Meanwhile, although the Hubble radius changes dramatically in the process --- $H$ vanishes momentarily at the bounce by definition --- 
 its magnitude at the beginning and end of the NEC-violating phase is nevertheless essentially the same, as shown with explicit bouncing solutions~\cite{us}.
Ignoring the details of this intervening phase, therefore, we can safely assume that $a$ and $|H|$ both match continuously
from the exit of the ekpyrotic phase to the onset of the radiation-dominated expanding phase, so that 
\be
\Omega_k^{({\rm ek-end})}\sim \Omega_k^{({\rm reheat})}\,.
\ee

Therefore, just as in inflation, starting with $\Omega_k^{({\rm ek-beg})}\sim \cO(1)$ at the onset of the ekpyrotic phase, one obtains an acceptably large radius of curvature by the present time
provided $N_{\rm ek} \;\gsim\; N_{\rm rad}$. In this precise sense the ekpyrotic scenario solves the flatness problem. 

The condition $N_{\rm ek} \;\gsim\; N_{\rm rad}$ implies that our entire observable universe is initially well within the proto-ekpyrotic patch.
Therefore, the assumption that the initial Hubble radius is reasonably smooth guarantees a high degree of homogeneity for our observable patch.

\subsection{Observational Constraints on Model Parameters}
\label{numbers}

We have seen that inflation and ekpyrosis are equally successful at addressing the well-known problems of standard big bang cosmology.
In particular, for the flatness problem, both allow for the natural choice $\Omega_k\sim \cO(1)$ initially.
Of course, in absolute terms the initial radius of curvature is much larger in ekpyrosis than in inflation. 
As stressed earlier, however, this is a consequence of strikingly different assumptions about initial conditions in the two models: inflation
proposes a hot beginning, corresponding to a microscopic proto-inflationary patch, whereas ekpyrotic theory prefers a cold and vacuous
initial state, corresponding to a macroscopically large Hubble radius. 

To make this point quantitative, in this subsection we estimate the initial Hubble radius in the ekpyrotic scenario for a range of reheating temperatures. First, using $N_{\rm ek} \;\gsim\; N_{\rm rad}$, we find from~(\ref{Nek}) that
\be
\left\vert H_{\rm ek-beg}\right\vert \;\lsim \; e^{-N_{\rm rad}} \left\vert H_{\rm ek-end} \right\vert = \frac{T_0}{T_{\rm reheat}}\left\vert H_{\rm ek-end} \right\vert  \,,
\label{Hekbeg}
\ee
where in the last step we have substituted~(\ref{Nrad}). As mentioned before and as shown in Fig.~\ref{timelines}, the net change in $|H|$ from the end of the ekpyrotic phase
until reheating in the expanding phase amounts to a factor of order unity, and thus $|H_{\rm ek-end}| \sim H_{\rm reheat}$. Moreover, from the Friedmann equation in a
radiation-dominated universe, we have $H_{\rm reheat}\sim T_{\rm reheat}^2/M_{\rm Pl}$. Substituting this into~(\ref{Hekbeg}), we obtain
\be
\left\vert H_{\rm ek-beg}\right\vert \;\lsim \; \frac{T_0T_{\rm reheat}}{M_{\rm Pl}} \,.
\label{Hekbeg2}
\ee
For $T_{\rm reheat}  = 10^{15}$~GeV, this gives $|H_{\rm ek-beg}| \;\lsim\; 10^{-6}$~eV, corresponding to an initial Hubble radius on the order of 1\;m.
For $T_{\rm reheat}  = 10^8$~GeV, we find $|H_{\rm ek-beg}| \;\lsim\; 10^{-13}$~eV, corresponding to a Hubble radius of the order of $10^4$~km. 

Note that the steepness of the ${\cal{V}}(\phi)$ factor in~\eqref{genpot} is characterized by the size of the fast-roll parameter $\epsilon$ defined in~\eqref{3.3}. In the more specific potential given in~\eqref{Vbew}, where the ${\cal{V}}(\phi)$ factor and the quadratic term are the same as in the two exponential case, we find that the steepness of the exponential is controlled by the constant $p=2\epsilon$. Restricting our discussion to~\eqref{Vbew}, we note for future reference that the parameter $p$ depends on the reheating temperature through the WMAP constraint on the amplitude of density perturbations--- see~(\ref{COBE}) and~(\ref{Delzeta}). Neglecting the model-dependent prefactor $\beta$ and again using $p=2\epsilon$, we can solve~(\ref{COBE}) and~(\ref{Delzeta}) for $p$:
\be
p \sim 10^{10} \left(\frac{H_{\rm reheat}}{M_{\rm Pl}}\right)^2\sim  10^{10} \left(\frac{T_{\rm reheat}}{M_{\rm Pl}}\right)^4\,.
\label{prel}
\ee
For instance, $T_{\rm reheat}  = 10^{15}$~GeV yields $p \sim 10^{-2}$. Notice that the fast-roll condition $p\ll 1$ combined with the WMAP amplitude puts an upper bound on the reheat temperature of $T_{\rm reheat} \;\lsim\; 10^{15}-10^{16}$~GeV. This is no different than in inflation since, as mentioned earlier,~(\ref{Delzeta}) is identical to its inflationary counterpart.

\section{Initial Conditions and Taming the Instability}
\label{taming}

In this section, we return to the tachyonic instability and study the associated initial conditions. We show that in order for the field trajectory to remain near $\chi\approx \chi_t=0$ sufficiently long to generate the appropriate range of scale-invariant perturbations, $\chi$ must be exponentially close to the top of the potential ridge at the onset of the ekpyrotic phase.
Were this to be an initial condition, it would imply that ekpyrotic theory requires incredible fine-tuning to achieve a realistic cosmology. 

Happily, as we will show below, this condition can easily and naturally be satisfied
without fine-tuning. This is accomplished by introducing a positive mass term for $\chi$, letting $\chi$  couple to light fermionic degrees of freedom and allowing for a pre-ekpyrotic phase where this positive mass dominates. Hence, in this early phase, the system will predominately oscillate around $\chi=0$.  For a wide range of initial conditions, the energy in these initial $\chi$ oscillations is efficiently 
transfered to the massless fermions, thereby bringing $\chi$
exponentially close to $\chi_t=0$ by the onset of the tachyonic phase.
Meanwhile, the energy density in the radiation produced quickly becomes subdominant
compared to that in $\phi$, thanks to the ekpyrotic attractor mechanism.

For ease of presentation, we will carry through our analysis not with the most general potential~\eqref{genpot} but, rather, with the more restricted potential~\eqref{Vbew}. We do this only for convenience. Our results easily extend to the general case~\eqref{genpot}. Recall that an essential 
feature of~\eqref{Vbew} is that the steepness of the pure exponential potential in $\phi$ is determined by the parameter $p \ll 1$.

\subsection{Constraints on Initial Conditions}
\label{consin}

To begin, let us carefully derive the constraints on the initial conditions associated with the instability in $\chi$ during the ekpyrotic phase. From the perturbation analysis in Sec.~\ref{pertrev}, it is clear that the tachyonic instability is crucial in generating a Harrison-Zeldovich spectrum for $\delta\chi$. This follows from the substitution of $V_{,\chi\chi} = -m_{\rm tachyon}^2 \approx -2/t^2$ into the fluctuation equation~\eqref{dchiflat}. Hence, we cannot remove this instability without spoiling scale invariance.
The main constraint associated with this instability is the following: the initial value of $\chi$ must be sufficiently close to $\chi_t=0$ so that the ekpyrotic phase driven by $\phi$ lasts long enough to generate perturbations over the desired range of modes. 

A necessary condition for this to be the case is that the energy density in $\chi$ be subdominant throughout the entire ekpyrotic phase. The total energy density during this phase is given by
\begin{equation}
\rho_{\rm total}=\rho_{\phi}+\rho_{\chi}\,,
\label{hope1}
\end{equation}
where we have defined
\begin{equation}
\rho_{\phi} = \frac{1}{2} {\dot{\phi}}^{2}+V_{\rm eff}(\phi)\,, \qquad 
\rho_{\chi}=\frac{1}{2}{\dot{\chi}}^{2} -\frac{1}{2}m_{\rm tachyon}^{2}\chi^{2}+{\cal{O}}(\chi^{3})\,,
\label{hope2}
\end{equation}
with $V_{\rm eff}(\phi)$ and $m_{\rm tachyon}$ given in~\eqref{hello} and~\eqref{mtachyon}, respectively. Hence, one must impose the constraint
\begin{equation}
|\rho_{\chi}| \; \lsim \; |\rho_{\phi}|
\label{hope2b}
\end{equation}
for all times during the ekpyrotic phase. In this case, we have
\begin{equation}
\rho_{\phi} \approx \rho_{\rm total} = 3H^{2}M_{\rm Pl}^{2}\,,
\label{hope3}
\end{equation}
where the last equality is just the Friedmann equation. Using expression~\eqref{hello2} and the fact that $p \ll 1$, it follows that one can ignore the Hubble term in the $\chi$ equation of motion. From~\eqref{5.2} and~\eqref{mtachyon}, this equation becomes
\begin{equation}
\ddot{\chi}-\frac{2}{t^{2}}\chi =0\,,
\label{hope4}
\end{equation}
whose growing-mode solution is 
\begin{equation}
\chi \sim (-t)^{-1}\,.
\label{hope5}
\end{equation}
It follows that the energy density in $\chi$ is 
\begin{equation}
\rho_{\chi} \approx -\frac{1}{4}m_{\rm tachyon}^{2} \chi^{2}\,.
\label{hope6}
\end{equation}

Returning to condition~\eqref{hope2}, we see that this constraint is strongest at $t_{\rm ek-end}$ marking the end of the ekpyrotic phase. At this time, it follows from~\eqref{hope2b},~\eqref{hope3} and~\eqref{hope6} that
\be
m_{\rm tachyon}(t_{\rm ek-end})\Delta\chi_{\rm ek-end} \;\lsim \; \left\vert H_{\rm ek-end}\right\vert M_{\rm Pl}\,,
\ee
where $\Delta\chi_{\rm ek-end}$ is the displacement of the field $\chi$ from the top of the ridge at the end of the ekpyrotic phase. Using~(\ref{mtachyon}) to substitute for $m_{\rm tachyon}$, we obtain
\be
\Delta\chi_{\rm ek-end} \;\lsim\; p\;M_{\rm Pl}\,.
\ee

Let us see what this entails for the $\chi$ displacement at the onset of the ekpyrotic phase, denoted by $\Delta\chi_{\rm ek-beg}$. Noting from the scaling solution~\eqref{5.2} and~(\ref{hope5}) that $\Delta\chi\sim |H|$, we find
\be
\Delta\chi_{\rm ek-beg} \;\lsim\; \left(\frac{H_{\rm ek-beg}}{H_{\rm ek-end}}\right)p\;M_{\rm Pl} 
= e^{-N_{\rm ek}} p\; M_{\rm Pl}\,,
\label{dchiek}
\ee
where in the last step we have substituted~(\ref{Nek}). 
In other words, $\chi$ must be exponentially close to $\chi_t=0$ at $t_{\rm ek-beg}$ in order for the ekpyrotic phase to last sufficiently long. 

We can also express this bound as a constraint on the initial energy density in $\chi$ at the beginning of the ekpyrotic phase:
\bea
\nonumber
\frac{\rho_\chi^{({\rm ek-beg})}}{\rho_\phi^{({\rm ek-beg})}} &\sim& \frac{m_{\rm tachyon}^2\left(t_{\rm ek-beg}\right)\Delta\chi^2_{\rm ek-beg}}{H_{\rm ek-beg}^2M_{\rm Pl}^2} \\
& \lsim & \; \frac{p^2m_{\rm tachyon}^2(t_{\rm ek-beg})}{H_{\rm ek-end}^2}
 \sim \left(\frac{H_{\rm ek-beg}}{H_{\rm ek-end}}\right)^2 = e^{-2N_{\rm ek}}\,.
\label{rhochi}
\eea
Thus the initial $\chi$ energy density has to be exponentially suppressed with respect to that of $\phi$ at the onset of the ekpyrotic phase. 

We conclude from~\eqref{dchiek} and~\eqref{rhochi} that the value of  $\chi$ at $t_{\rm ek-beg}$ has to be exponentially close to the top of the tachyonic ridge in order to generate the necessary e-foldings of scale-invariant perturbations. As an initial condition, this introduces a fine-tuning into ekpyrotic theory. However, as mentioned in the introduction to this section, the same result can be obtained naturally, without fine-tuning. We now show how this can be achieved.


\subsection{Stabilization Mechanism}
\label{stable}


In this subsection, we argue that the bounds~\eqref{dchiek} and~\eqref{rhochi} are naturally satisfied if the ekpyrotic phase, during which $\chi$ is tachyonic, is preceeded by a phase in which $\chi$ has positive mass squared and couples to light fermions. This can be achieved, for example, by adding a mass term $m_\chi^2(\phi)\chi^2$ to~(\ref{Vbew}) where, in general, $m_{\chi}$ 
depends on $\phi$. The potential  then is given by
\be
V(\phi,\chi) = -V_0e^{-\sqrt{\frac{2}{p}}\;\phi/M_{\rm Pl}}\left(1+\frac{1}{pM_{\rm Pl}^2}\chi^2+
\ldots \right) + \frac{1}{2}m_\chi^2(\phi)\chi^2 \,.
\label{4.7}
\ee
The new mass parameter $m_{\chi}(\phi)$ is chosen so that effective mass term for $\chi$,
\be
m_{\rm eff}^2(\phi) = -m_{\rm tachyon}^2(\phi) + m_\chi^2(\phi)\,,
\ee
is positive at early times but becomes tachyonic for sufficiently small 
values of $\phi$. This can be the case even for constant $m_\chi$, since $m_{\rm tachyon}$ grows in time as $\phi$ evolves down the exponential potential. However, this transition is facilitated by choosing $m_{\chi}(\phi)$ to be a decreasing function of $\phi$. The end of the pre-ekpyrotic phase and the beginning of the ekpyrotic phase occur by definition at time $t_{\rm ek-beg}$ when $\chi$ becomes tachyonic:
\be
m_\chi(t_{\rm ek-beg}) \;\lsim\; m_{\rm tachyon}(t_{\rm ek-beg}) \sim \frac{|H_{\rm ek-beg}|}{p}\,. 
\label{m_6}
\ee
After this time the dynamics is dominated by the ekpyrotic potential~\eqref{Vbew}. A typical potential of the form~\eqref{4.7} is shown in Fig.~\ref{2fieldpotstab}.

In order to generate the desired number of e-folds of perturbations,  $m_\chi$ must become subdominant at a sufficiently early time. Otherwise, the tachyonic phase will be triggered too late in the evolution of the universe, resulting in too short a range of scale-invariant perturbations. Assuming that the largest observable mode exits the horizon at $t_{\rm ek-beg}$, a conservative assumption, we can use~\eqref{m_6} to put an upper bound on the allowed value of $m_{\chi}$ at $t_{\rm ek-beg}$. 
Expressing the right-hand side of~\eqref{m_6} in terms of the reheating temperature by substituting for $H_{\rm ek-beg}$ and $p$ through~(\ref{Hekbeg2}) and~(\ref{prel}) respectively, we find that
\be
m_\chi (t_{\rm ek-beg}) \;\lsim\; 10^{-10}\frac{T_0}{M_{\rm Pl}}\left(\frac{M_{\rm Pl}}{T_{\rm reheat}}\right)^3\;M_{\rm Pl} \approx 10^{-40}\left(\frac{M_{\rm Pl}}{T_{\rm reheat}}\right)^3\;M_{\rm Pl}  \,.
\label{m_62}
\ee
In the last step, we have used the fact that the microwave background temperature is $T_{0} \approx 
2.7\;K\approx 10^{-30} M_{\rm Pl}$. For example, $T_{\rm reheat}  = 10^{15}$~GeV gives $m_\chi (t_{\rm ek-beg})\;\lsim\; 10^{-4}$~eV, whereas for $T_{\rm reheat}  = 10^{8}$~GeV the bound greatly weakens to $m_\chi (t_{\rm ek-beg})\;\lsim\; 10^8$~GeV. Finally, for $T_{\rm reheat}  \;\lsim\; 10^5$~GeV, we get a trivial bound of $m_\chi (t_{\rm ek-beg})\;\lsim\; 0.1\;M_{\rm Pl}$.

The pre-ekpyrotic phase occurs during the time interval $t_i \leq t \leq t_{\rm ek-beg}$, where $t_{i}$ is some as yet unspecified initial time. By definition, during this phase $\chi$ has positive mass squared given, for most of the epoch, by $m_{\rm eff}^{2} \approx m_{\chi}^{2}$. Given some initial 
displacement $\Delta\chi_{i}$, $\chi$ will undergo oscillations around the minimum of the potential at $\chi=0$. Since the universe is contracting, we cannot rely on Hubble friction to damp out these oscillations. On the contrary, they will instead get amplified by the usual blueshift effect. However, if we assume that $\chi$ is coupled to light fermionic degrees of freedom, then its coherent oscillations will decay into these particles, thereby stabilizing $\chi$.
Let $\Psi$ be a typical light fermion and assume that it couples to $\chi$ as $\chi \bar{\Psi} \Psi$. Then the rate of decay is given as usual by~\cite{kolbturner}
\be
\Gamma \sim m_\chi\,.
\ee
If $\Gamma\gg |H|$, the decay is rapid on a Hubble time and the $\chi$ evolution is overdamped. That is,  the field settles at its minimum well before the blueshift effect can amplify the oscillations.
This is easily satisfied here if we choose $m_\chi$ at $t_{\rm ek-beg}$ to saturate the bound~(\ref{m_6}) since, in this case, we have
\be
m_\chi (t_{\rm ek-beg})\sim \frac{|H_{\rm ek-beg}|}{p}\gg |H_{\rm ek-beg}|\;\gsim\; |H_i|\,,
\label{dog1aa}
\ee
where $H_i$ is the initial Hubble parameter. Since $m_{\chi}$ is either constant or increases at earlier times, it follows that $\Gamma \gg |H|$ at any time during the pre-ekpyrotic phase.

Let us now calculate the initial conditions at the beginning of the pre-ekpyrotic phase that will produce the exponentially suppressed displacement~\eqref{dchiek} and ratio of densities~\eqref{rhochi} 
at $t_{\rm ek-beg}$. As in the ekpyrotic phase, we will assume for concreteness that
\begin{equation}
|\rho_{\chi}| \;\lsim\; |\rho_{\phi}|\,,
\label{dog1}
\end{equation}
so that the evolution is dominated by $\phi$ during the entire pre-ekpyrotic epoch. It will soon become obvious that our stabilization mechanism is sufficiently effective
to allow for larger initial $\chi$ component. Of course the mechanism eventually fails if $\rho_\chi$ is initially orders of magnitude larger than $\rho_\phi$, but these are
clearly fine-tuned and unnatural initial conditions.

To simplify the calculation, we henceforth take $m_\chi$ to be independent of $\phi$. Moreover, we assume that $m_\chi$ saturates~(\ref{dog1aa}),
\begin{equation}
m_\chi \sim m_{\rm tachyon}(t_{\rm ek-beg})
\label{dog1a}
\end{equation}
so that $\Gamma\gg |H|$. In this regime, we can ignore any blueshift effect. It follows that from the earlier discussion that
\begin{equation}
\rho_{\phi} \approx 3H^{2}M_{\rm Pl}^{2}\;, \qquad |\rho_{\chi}| \approx \frac{1}{4}m_{\chi}^{2}\chi^2.
\label{dog2}
\end{equation}
Putting these densities into~\eqref{dog1} and evaluating the inequality at $t_{i}$ gives
\begin{equation}
m_{\chi}(t_{i})\Delta\chi_{i} \;\lsim \;\left\vert H_{i}\right\vert M_{\rm Pl}\;,
\label{dog3}
\end{equation}
where $\Delta\chi_{i}$ is the displacement of field $\chi$ at the beginning of the pre-ekpyrotic phase.
Using~\eqref{dog1aa} and~\eqref{dog1a}, we find that the initial displacement must satisfy
\begin{equation}
\Delta\chi_{i}\; \lsim \left(\frac{H_{i}}{H_{\rm ek-beg}}\right)pM_{\rm Pl}\,.
\label{dog4}
\end{equation}
This consistency condition ensures that $\chi$ is subdominant throughout the pre-ekpyrotic phase.

To determine the allowed values of $\Delta\chi_{i}$, one must evaluate the ratio $H_{i}/H_{\rm ek-beg}$.
This is most easily done by considering the ratio of the energy densities. From the above discussion it follows that the energy density in $\chi$ decays as $\rho_\chi \propto e^{-\Gamma t}\sim e^{-m_\chi t}$. Thus, starting with some energy density $\rho_\chi^{(i)}$ at $t_i$, by the onset of the ekpyrotic phase this has decayed to a value $\rho_\chi^{({\rm ek-beg})}$ given by
\be
\frac{\rho_\chi^{({\rm ek-beg})}}{\rho_\chi^{(i)}} \approx e^{-m_\chi (t_{\rm ek-beg}-t_i)}\approx  e^{-m_\chi (-t_i)} \approx e^{-\sqrt{2}|H_{\rm ek-beg}|(-t_i)/p}  \,,
\label{expsup}
\ee
where in the second step we have assumed, for simplicity, that the pre-ekpyrotic phase is long compared to $t_{\rm ek-beg}$; that is, $-t_i\gg -t_{\rm ek-beg}$. Moreover we have substituted $m_\chi \sim m_{\rm tachyon}(t_{\rm ek-beg})\approx \sqrt{2}|H_{\rm ek-beg}|/p$ from~(\ref{mtachyon}). 

It is convenient to rewrite this in terms of the initial Hubble parameter $H_i$. Since $\rho_\chi^{(i)} \;\lsim \; \rho_\phi^{(i)}$ by assumption, the cosmological evolution
is dominated by $\phi$ throughout the pre-ekpyrotic phase. Therefore, we can substitute the relation $H_i = p/t_i$ given in~\eqref{5.2}:
\be
\frac{\rho_\chi^{({\rm ek-beg})}}{\rho_\chi^{(i)}}  \approx e^{-\sqrt{2}H_{\rm ek-beg}/H_{i}}\,,
\ee
and thus
\be
\frac{\rho_\chi^{({\rm ek-beg})}}{\rho_\phi^{({\rm ek-beg})}} \approx \frac{\rho_\chi^{(i)}}{\rho_\phi^{({\rm ek-beg})}} e^{-\sqrt{2}H_{\rm ek-beg}/H_{i}}\ <  e^{-\sqrt{2}H_{\rm ek-beg}/H_{i}}\,.
\label{rhobd}
\ee
The last inequality follows from $\rho_\phi^{({\rm ek-beg})} > \rho_\phi^{(i)}$ (due to blueshift between $t_i$ and $t_{\rm ek-beg}$) as well as $\rho_\phi^{(i)} \;\gsim\; \rho_\chi^{(i)}$ (by assumption). Comparing this against the bound given in~(\ref{rhochi}), we see that it  is satisfied if 
\be
\frac{H_{\rm ek-beg}}{H_i} \;\gsim \; \sqrt{2}N_{\rm ek}\,.
\label{bdcase1}
\ee
Putting this into~\eqref{dog4}, we get an expression for the displacement at the beginning of the pre-ekpyrotic phase given by
\begin{equation}
\Delta\chi_{i}\;\lsim\;\frac{1}{\sqrt{2}N_{\rm ek}}\;pM_{\rm Pl}\,.
\label{dog6}
\end{equation}

To get a feel for what this expression entails, recall from Section 4 that $p$ and $N_{\rm ek}$ depend on the reheat temperature. For example, for $T_{\rm reheat}=10^{15} $ GeV, we found that $p\;\sim\;10^{-2}$ and $N_{\rm ek} \;\gsim\; 60$. It follows that for this reheat temperature the displacement at the beginning of the pre-ekpyrotic phase is given by
\begin{equation}
\Delta\chi_{i}\;\lsim\; 10^{-4} M_{\rm Pl}\,,
\label{dog7}
\end{equation}
representing a wide range of natural initial conditions. 

It is also useful to consider the ratio of the displacement of $\chi$ at $t_{i}$ to the exponentially fine-tuned value of $\chi$ required at $t_{\rm ek-beg}$. Using~\eqref{dchiek} and~\eqref{dog6}, we find that
\begin{equation}
\frac{\Delta\chi_{i}}{\Delta\chi_{\rm ek-beg}}\; \sim\; \frac{e^{N_{\rm ek}}}{\sqrt{2} N_{\rm ek}}\,.
\label{dog8}
\end{equation}
Clearly our mechanism naturally removes the exponential fine-tuning problem, as claimed.

The remaining issue is to understand the physical implication of choosing the Hubble parameters at $t_{\rm ek-beg}$ and $t_{i}$ to satisfy the ratio~\eqref{bdcase1}. Using the fact that during the $\phi$-dominated pre-ekpyrotic phase $H=p/t$, it follows from~\eqref{bdcase1} that
\begin{equation}
\frac{t_{i}}{t_{\rm ek-beg}} \; \gsim \; \sqrt{2}N_{\rm ek}\,,
\label{dog9}
\end{equation}
a natural and physically reasonable assumption. For example, for the reheat temperature $T_{\rm reheat}=10^{15} $ GeV this ratio becomes
\begin{equation}
\frac{t_{i}}{t_{\rm ek-beg}} \; \gsim \; 10^{2}\,.
\label{dog10}
\end{equation}

We conclude that in order for $\chi$ to be close enough to the top of the tachyonic ridge by the onset of the ekpyrotic phase to ensure a sufficient number of e-foldings, it suffices to start out at
$t_i$ with the large, natural value of $\chi$ given by~\eqref{dog6}. This will be the case as long as 
the pre-ekpyrotic phase begins sufficiently before the onset of ekpyrosis, the precise relationship being specified in~\eqref{dog9}.

\section{Graceful Exit and Bouncing}
\label{end}

Thus far, we have discussed 1) the pre-ekpyrotic phase in which $\chi$ gets stabilized near $\chi_t=0$
and 2) the ekpyrotic phase in which the entropy perturbation acquires a nearly
scale-invariant spectrum. In this section,  we turn to the final two phases required for a realistic cosmology, that is, 3) the exit from the ekpyrotic phase and 4) the bounce from contraction to expansion enabled by a NEC-violating ghost condensate. We begin with a general discussion of these last two phases, in Secs.~\ref{convert} and~\ref{bounce} respectively, before turning to a specific example in Sec.~\ref{example}.


\subsection{Converting Entropy into Curvature Perturbations}
\label{convert}


The ekpyrotic phase described by the scaling solution~(\ref{5.2}) must eventually come to an end if the universe is to undergo a smooth bounce and enter the hot big bang expanding phase.
Before turning to the bounce, however, we first discuss a key element of New Ekpyrotic Cosmology which also occurs during the exit from ekpyrosis, namely the conversion of the scale-invariant entropy spectrum into the adiabatic mode. Even before the bounce, the exit from the ekpyrotic phase can be used to imprint the scale-invariant entropy spectrum onto the adiabatic mode, described by $\zeta$, the curvature perturbation on uniform-density hypersurfaces. 

\begin{figure}[ht]
\centering
\includegraphics[width=90mm]{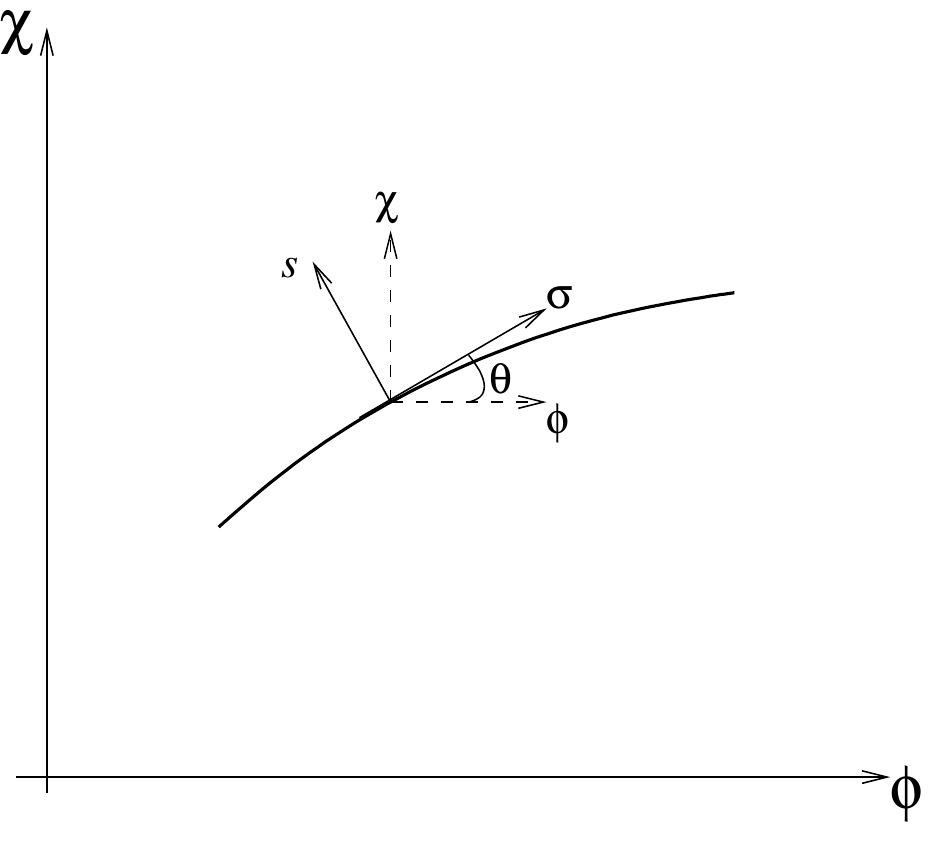}
\caption{Adiabatic ($\sigma$) and entropy ($s$) directions at a given point along the field trajectory in $(\phi,\chi)$ space.}
\label{rotate}
\end{figure}

In general, it is convenient to consider new field variables, $\sigma$ and $s$, defined with respect to the field trajectory. As shown in Fig.~\ref{rotate}, 
the adiabatic field velocity, $\dot{\sigma}$, is proportional to the vector tangent to the curve, while the entropy field velocity, $\dot{s}$, is
proportional to the normal vector. In other words, their decomposition in terms of $\dot{\phi}$ and $\dot{\chi}$ is given by
\bea
\nonumber
\dot{\sigma} &=& \cos\theta\, \dot{\phi} + \sin\theta\,\dot{\chi} \\
\dot{s} &=& -\sin\theta\,\dot{\phi} + \cos\theta\, \dot{\chi} \,,
\label{deltafield}
\eea
where
\be
\tan \theta = \frac{\dot{\chi}}{\dot{\phi}}\,.
\label{ttheta}
\ee

During the ekpyrotic phase, the background field motion is along $\phi$, and thus $\theta = 0$. Evidently, it follows from~(\ref{deltafield}) that in this phase
$\sigma$ and $s$ coincide with $\phi$ and $\chi$, respectively. Once we exit the ekpyrotic phase, this is of course no longer true. In general, the ($\phi,\chi$)
equations of motion given by~(\ref{eomsphichi}) imply the following evolution equation for the adiabatic field
\be
\ddot{\sigma} + 3H\dot{\sigma} = -V_{,\sigma}\,,
\label{sigmaeom}
\ee
where $V_{,\sigma}\equiv \cos\theta V_{,\phi} + \sin\theta V_{,\chi}$. This result will come in handy later on in the discussion.

The relation between the adiabatic and entropy fluctuations to $\delta\phi$ and $\delta\chi$ follows immediately from~(\ref{deltafield}):
\bea
\nonumber
\delta\sigma &=& \cos\theta\, \delta\phi + \sin\theta\,\delta\chi \\
\delta s &=& -\sin\theta\,\delta\phi + \cos\theta\, \delta\chi \,.
\label{deltafieldpert}
\eea
Using the standard energy and momentum constraints, the entropy perturbation satisfies~\cite{gordon}
\be
\ddot{\delta s_k} + 3H\dot{\delta s_k} + \left(\frac{k^2}{a^2}+V_{,ss}+
3\dot{\theta}^2\right)\delta s_k =
4M_{\rm Pl}^2\frac{\dot{\theta}}{\dot{\sigma}}\frac{k^2}{a^2}\Psi_k\,,
\label{ds}
\ee
where $\Psi$ is the curvature perturbation in Newtonian gauge, while
$V_{,ss}  = \cos^2\theta\, V_{,\chi\chi} + \sin^2\theta\, V_{,\phi\phi}$
is the curvature of the potential orthogonal to the field trajectory.

Since $\theta = 0$ during the ekpyrotic phase, as mentioned earlier, we  have $\delta s = \delta \chi$,
so that fluctuations in $\chi$ indeed correspond to entropy perturbations. As a quick consistency check,
since $\dot{\theta} = 0$ and $V_{,ss}  = V_{,\chi\chi}$, we find that~(\ref{ds}) indeed reduces to~(\ref{6.4}). 
Note, however, that the identification of $\delta\chi$ with the entropy perturbation only
holds during the ekpyrotic phase; that is, as long as the field
trajectory is along the $\phi$ direction. We will shortly introduce a feature in the potential that
causes a sharp turn in the field trajectory at the end of the ekpyrotic phase, thereby imprinting
a scale-invariant spectrum on $\zeta$, which is proportional to the adiabatic fluctuation $\delta\sigma$. Subsequently the classical field evolution is partially along $\chi$, and, evidently,
fluctuations in the latter are no longer identified with the entropy perturbation.

Let us therefore turn to the conversion of the entropy fluctuation to the curvature perturbation spectrum. The latter is sourced by the entropy perturbation as follows~\cite{gordon}
\begin{equation}
\dot{\zeta_k} = \frac{H}{\dot{H}}\frac{k^2}{a^2}\Psi_k +  \frac{2H}{\dot{\sigma}}\dot{\theta}\delta s_k\,.
\label{5.3}
\end{equation}
The first term proportional to $k^2$ is seemingly negligible at long wavelengths. As pointed out in~\cite{wands2},
however, this is not necessarily the case during the exit from the ekpyrotic phase. For our purposes we will neglect this term altogether for simplicity.

Clearly, to endow $\zeta$ with a scale-invariant contribution from $\delta s$, the field trajectory must undergo some turn
at the end of ekpyrosis, which we assume is sharp and therefore rapid on a Hubble time. In~\cite{us}, this was achieved by introducing a minimum followed by a sharp rise in the respective potentials for $\phi_1$ and $\phi_2$. In the simplified $\phi,\chi$ framework studied here, this would correspond to adding new terms in the potential that depend on specific linear combinations of $\phi$ and $\chi$.
A much simpler option suggests itself, however, namely to introduce some term which pushes $\chi$ away from $\chi_{t}=0$ at the end of the ekpyrotic phase.

To be precise, we want to modify potential~\eqref{Vbew} so that $\chi_t=0$ is only an approximate solution to the equations of motion. One can, for example, add a term to the potential
which is approximately linear in $\chi$ near $\chi=0$ and therefore prevents the latter from being an extremum of the potential. This linear term must be a function of $\phi$
so that it is negligible during the ekpyrotic phase but eventually becomes important after a sufficient number of e-foldings is achieved. At this point it tips $\chi$ away from the
tachyonic ridge marking the end of the ekpyrotic phase. This is sketched in Fig.~\ref{2fieldpotend}.

To illustrate this, let us describe the ekpyrotic phase by the simple potential~\eqref{4.7} and add a correction of the form
\begin{equation}
\Delta V = U_0 e^{-\sqrt{\frac{2}{p^{\prime}}}\phi/M_{\rm{Pl}}}
\left(1+\frac{1}{p^{\prime}M^2_{\rm{Pl}}}(\chi-\chi_0)^2 \right)\,.
\label{egpot}
\end{equation}
Clearly, this has non-vanishing slope at $\chi=0$. For $p'\ll p$, the exponential prefactor makes this correction vanishingly small
during the ekpyrotic phase, and therefore $\chi_t \approx 0$ is still a valid solution then. Once $\phi$ becomes sufficiently small, however, the correction~\eqref{egpot}
becomes important and pushes $\chi$ away from the $\chi_t=0$ tachyonic point. This example will be studied in detail in Sec.~\ref{example}.

Coming back to general considerations, we assume that the driving force is sufficiently strong to generate a fast-roll of $\chi$ away from the tachyonic point,
corresponding to a sharp turn in the field trajectory. To make this more precise, we require that the driving term $V_{,\chi}$ in the equation of motion
for $\chi$ --- see~(\ref{eomsphichi}) --- is comparable to the Hubble damping term:
\be
|V_{,\chi}(\chi=0)|\;\gsim\; M_{\rm Pl}H^2\,.
\label{condVchi}
\ee
Equivalently, this condition amounts to requiring the slow-roll parameter $\epsilon_\chi \equiv V_{,\chi}^2/H^4M_{\rm Pl}^2$ to be comparable or greater than
unity. For example, for the specific correction given in~(\ref{egpot}), this condition is satisfied for sufficiently small values of $\phi$. We henceforth denote by $\phi_{\rm ek-end}$
the value of $\phi$ when~(\ref{condVchi}) is fulfilled. This notation is consistent with Fig.~\ref{timelines} since the rolling of $\chi$ away from the
tachyonic ridge marks the end of the ekpyrotic phase for all practical purposes. In particular, fluctuation modes produced thereafter are no longer scale-invariant.

In the limit of fast-roll, we can treat the turn in the field trajectory as 
instantaneous on a Hubble time. This amounts to approximating $\dot{\theta}$ 
with a delta function and treating $H$ as constant. Hence
\begin{equation}
\dot{\zeta_k} 
\approx \frac{2H_{\rm ek-end}}{\dot{\sigma}}\Delta\theta\;  \delta(t-t_{\rm ek-end})\;\delta s_k\,,
\label{5.3b}
\end{equation}
where $\Delta\theta$ is the total change in angle during the exit. Of course, smoother exit dynamics could do equally well,
but this rapid-exit approximation makes it possible to proceed analytically.

As argued in~\cite{us}, $\dot{\sigma}$ is constant during the instantaneous exit from the ekpyrotic phase.
This follows by inspection of the equation of motion for $\sigma$ given in~(\ref{sigmaeom}). If $\dot{\sigma}$ is not constant during the transition, then the $\ddot{\sigma}$ term will generate a delta-function contribution.
But this cannot be compensated by any other term in the equation. The $3H\dot{\sigma}$ evidently cannot do the trick, since
$H$ is constant during the transition by assumption. Similarly, a sudden change in field direction can at most yield a jump in $V_{,\sigma}$,
but not a delta-function discontinuity. Thus $\dot{\sigma}$ is continuous, and we can substitute its value at the end of the ekpyrotic phase.
Since $\dot{\sigma} = \dot{\phi}$ during this phase, we read off from~\eqref{5.2}  
\be
\dot{\sigma} = \frac{2\sqrt{\epsilon}M_{\rm Pl}}{t_{\rm ek-end}} = \frac{\left\vert H_{\rm ek-end}\right\vert}{\sqrt{\epsilon}}M_{\rm Pl}\,, 
\ee
where we have used $p=2\epsilon$.

Similarly $\delta s$ changes by at most a factor of unity during the exit, and we refer the reader to~\cite{us} for a detailed
argument. Thus we approximate $\delta s$ to match continuously as well, for simplicity, and can substitute for $\delta s$ its value
at the exit from the scaling solution. Since $\delta s= \delta \chi$ up to the exit, this can be read off from~\eqref{5.2} and~\eqref{again}, again using $p=2\epsilon$: 
\be
k^{3/2}\delta s_k = k^{3/2} \frac{\left\vert H_{\rm ek-end}\right\vert}{2^{3/2}\epsilon}\,.
\ee

Substituting the above expressions for $\dot{\sigma}$ and $\delta s$, we can integrate~(\ref{5.3b}) and obtain
\be
\Delta_\zeta^{1/2} = k^{3/2}\zeta_k \approx \Delta\theta \frac{\left\vert H_{\rm ek-end}\right\vert}{\sqrt{2\epsilon}M_{\rm Pl}} \approx \Delta\theta \frac{H_{\rm reheat}}{\sqrt{2\epsilon}M_{\rm Pl}}\,,
\label{5.6}
\ee
where we have used $|H_{\rm ek-end}|\sim H_{\rm reheat}$ as indicated in Fig.~\ref{timelines} and discussed earlier. Comparison with~(\ref{Delzeta}) shows that the model-dependent
factor $\beta$ is identified with $\Delta\theta$ in our sharp turn approximation, thereby confirming the discussion below~(\ref{Delzeta}). It is desirable to have $\Delta\theta\sim \cO(1)$, for otherwise
one needs even smaller values of $\epsilon$ at fixed reheating temperature to maintain the WMAP normalization~(\ref{COBE}). In the next subsection we will discuss the conditions for which
$\Delta\theta\sim \cO(1)$.

To summarize, the end of the ekpyrotic phase is set by a term in the potential which pushes $\chi$ away
from the $\chi_{t}=0$ tachyonic point. This term is assumed irrelevant during the ekpyrotic phase, so that $\chi\approx 0$ is a good
approximate solution, but eventually becomes important and triggers the end of ekpyrosis. This sudden motion in $\chi$ corresponds
to a turn in the field trajectory, which in turn imprints the scale-invariant entropy perturbation spectrum onto the curvature perturbation.

Instead of dialing the end of the ekpyrotic phase with additional terms in the potential, one could alternatively exploit the tachyonic instability in $\chi$, as studied in~\cite{wands1, wands2}.
However, as the discussion of Sec.~\ref{consin} most emphatically underscores, to generate the desired number of e-foldings in this case requires an exponential fine-tuning in the initial $\chi$ displacement --- see, for example,~(\ref{dchiek}). If the initial displacement is larger, then the ekpyrotic phase will be too short, resulting in an unacceptably narrow spectral range of scale-invariant perturbations. 
If it is smaller, on the other hand, then the ekpyrotic phase will last for too long. That is, by the time the entropy perturbation is converted onto $\zeta$, the Hubble parameter will be too large in magnitude to yield an acceptable perturbation amplitude. In other words, the WMAP constraint on the amplitude of perturbations is tied in this latter case to a tuning of initial conditions.

On the other hand, adding correction terms to the potential, as we do here, leads to robust predictions for the perturbation spectrum for a broad range of initial conditions. Moreover, we know that corrections to~(\ref{4.7}) are necessary for independent reasons. For instance, the NEC-violating ghost condensation phase requires a nearly flat and positive potential. Thus the negative, steep exponential
potential of the ekpyrotic phase must eventually have a minimum and rise to positive values. What we are advocating is that these corrections will generically push $\chi$ away from $\chi=0$, thereby endowing $\zeta$ with a scale-invariant spectrum. We will illustrate this in detail in Sec.~\ref{example} with the correction~(\ref{egpot}), showing how it leads to a scale-invariant curvature perturbation and how it can be matched smoothly to the NEC-violating ghost condensate phase.


\subsection{A Bouncing Scenario}
\label{bounce}


In this subsection, we complete the scenario by discussing the dynamics of
a non-singular bounce along the lines of~\cite{us}. This is achieved by
supplementing the ekpyrotic phase with a phase in which the energy density is dominated
by that of a ghost condensate~\cite{Arkani}. The ghost condensate can violate the null energy condition~\cite{Paolo},
a necessary ingredient to produce a bounce, without introducing ghost-like instabilities. 
The analysis closely parallels that of~\cite{us}, and thus we will be brief, referring the reader to our earlier paper for details.
A key difference, however, is that here it suffices to have only one field entering the ghost condensate phase, namely
the ekpyrotic field $\phi$; meanwhile $\chi$ is a standard scalar field with canonical kinetic term throughout.

\vspace{0.5cm}
\noindent {\it i) Rolling of $\chi$ towards minimum}
\vspace{0.2cm}

Before entering the ghost condensate phase, we assume that $\chi$ gets stabilized. 
As described in the previous subsection, the production of scale-invariant modes
comes to a halt once corrections to the ekpyrotic potential drive $\chi$ away from the tachyonic point. 
Such corrections are assumed negligible at early times but become important for $\phi$ smaller than some critical value $\phi_{\rm ek-end}$, where 
the subscript indicates that this marks the end of the ekpyrotic phase for all practical purposes.

Subsequently we assume that $\chi$ rolls towards a stable minimum, denoted by $\chi_{\rm min}$. For instance, our fiducial correction term given in~(\ref{egpot}) generates a minimum at $\chi_{\rm min}\approx \chi_0$. Note that the fermionic degrees of freedom which stabilized the field in the pre-ekpyrotic phase now become heavy as $\chi$ acquires a non-zero expectation value. Thus they become irrelevant and cannot be relied upon  to stabilize $\chi$ in this post-ekpyrotic phase. Instead $\chi$ rolls down and undergoes undamped harmonic oscillations around $\chi_{\rm min}$. 

The kinetic energy acquired by $\chi$ as it rolls away from the ridge is of course just given by the difference in potential energy between $\chi=0$ and $\chi=\chi_{\rm min}$:
\be
\rho_\chi (t_{\rm ek-end}) =  \frac{1}{2}\dot{\chi}^2(t_{\rm ek-end}) = V\left(\phi_{\rm ek-end}, 0\right) - V\left(\phi_{\rm ek-end},\chi_{\rm min}\right)\,.
\label{rhorad}
\ee
If we assume for simplicity that $\chi$ is massive compared to Hubble,
\be
V_{,\chi\chi}\left(\chi_{\rm min}\right)  \gg H^2\,,
\ee
then the field undergoes many oscillations in a Hubble time. Consequently, its energy density, averaged over many oscillations, blueshifts as dust: 
\be
\rho_\chi\sim a^{-3}\,. 
\label{rhoa3}
\ee
Although this grows in time as the universe contracts, nevertheless $\rho_\chi$ quickly becomes negligible compared to the energy density in $\phi$, which keeps rolling down
its steep quasi-exponential potential. In other words, thanks to the ekpyrotic attractor mechanism, we can tolerate
$\rho_\chi \gg \rho_\phi$ at $t_{\rm ek-end}$ since $\phi$ will dominate again within a few e-folds of contraction.

These considerations impact on the allowed range of $\Delta\theta$, the change in angle in the field trajectory as we exit the ekpyrotic phase.
As shown in~(\ref{5.6}), this parameter enters in the amplitude of density perturbations.  Here we argue that a large change in angle is indeed possible.

From the definition of $\theta$ in~(\ref{ttheta}), we see that $\Delta\theta \sim \cO(1)$ is achieved provided 
\be
\dot{\chi}^2(t_{\rm ek-end})\;\gsim\; \dot{\phi}^2(t_{\rm ek-end})\,.
\ee
Equivalently, since $\rho_\chi (t_{\rm ek-end})\approx \dot{\chi}^2(t_{\rm ek-end})/2$ from~(\ref{rhorad}) whereas
$\rho_\phi \approx p\; \dot{\phi}^2/2$ from~(\ref{5.2}), we need
\be
\rho_\chi \gg \rho_\phi\qquad {\rm at}\;\;t=t_{\rm ek-end}\,.
\ee
The discussion of the previous paragraph shows that such a large ratio of densities is allowed by the ekpyrotic attractor mechanism.
In other words, $\chi$ can dominate momentarily as it rolls off the ridge and starts oscillating, thereby generating a significant $\Delta\theta$.
Very soon, however, $\phi$ will take over as it keeps rolling down its steep negative exponential potential.

\begin{figure}[ht]
\centering
\includegraphics[width=100mm]{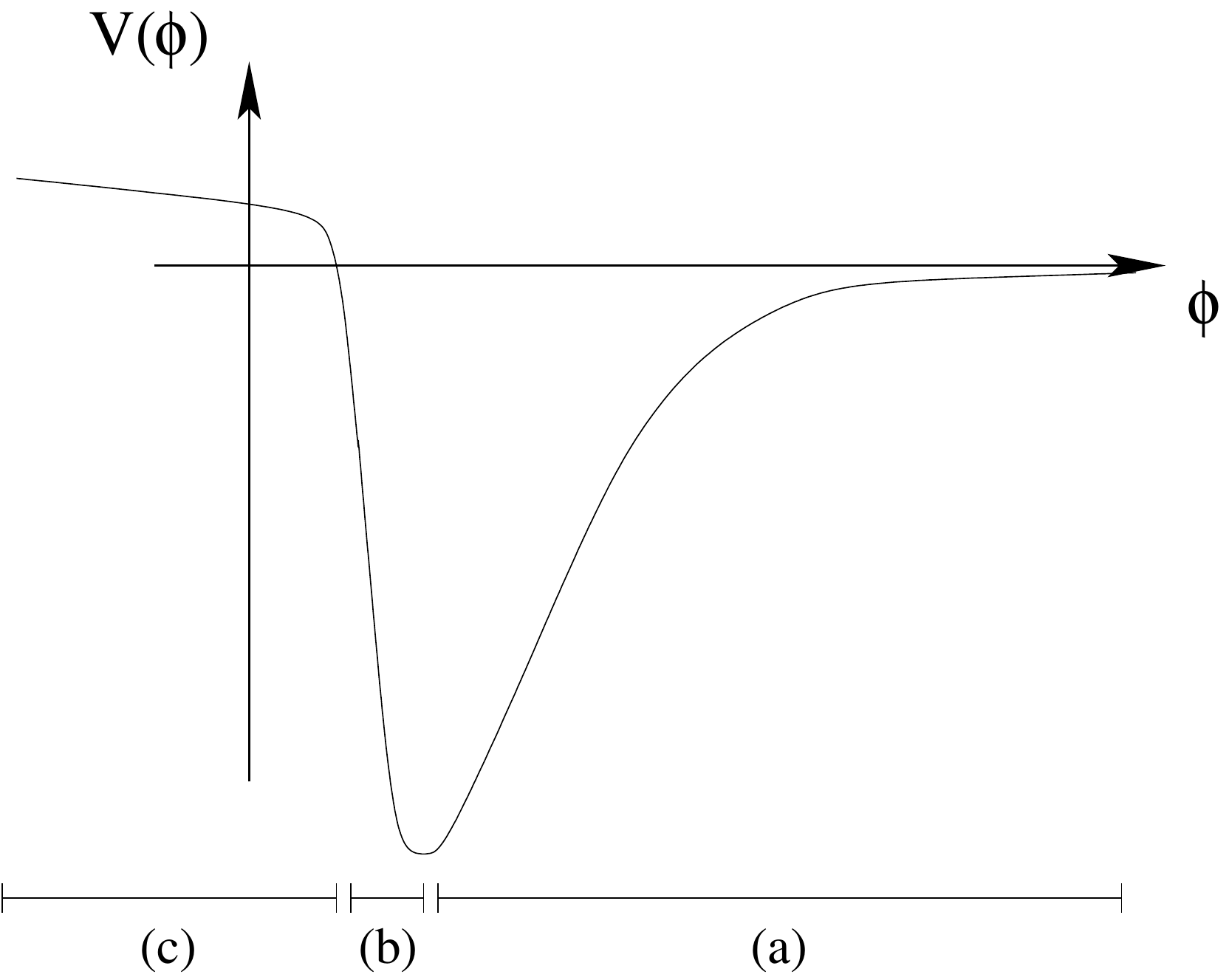}
\caption{Effective potential for $\phi$ at fixed $\chi$.}
\label{onefield}
\end{figure}

Once $\chi$ is subdominant, the dynamics effectively reduce to that of a single field $\phi$, and the story is virtually identical to single-field ekpyrosis with NEC-violating ghost condensate and non-singular bounce. The effective single-field potential is shown in Fig.~\ref{onefield}. Region a) describes a steep negative exponential potential corresponding to the ekpyrotic phase. 
A necessary condition to violate the NEC is that the potential becomes {\it positive}. To see this, recall that the onset of NEC-violation occurs when $\dot{H} = 0$. And since $M_{\rm Pl}^2\dot{H} = -\dot{\phi}^2/2$, this also corresponds to $\dot{\phi}^2 =0$. But then the Friedmann equation, $3H^2M_{\rm Pl}^2 = \dot{\phi}^2/2 +V $, immediately tells us that $V > 0$. 
Thus the field must reach a minimum in $V$ 
and rise to positive values (region b)). 
Such a minimum and sharp rise can be achieved with our ekpyrotic potential~(\ref{4.7}) plus the
correction~(\ref{egpot}). The region c) corresponds to the ghost condensation 
phase where the potential is approximately flat. 

\vspace{0.5cm}
\noindent {\it ii) Ghost condensate phase and NEC violation}
\vspace{0.2cm}

Our next consideration is the epoch of NEC violation, triggered by $\phi$ entering a ghost condensate
phase. This is achieved by invoking higher-derivative corrections  
to the kinetic term for $\phi$, which takes the form
\begin{equation}
{\cal{L}}= \sqrt{-g} M^{4}P(X),
\label{evgeny}
\end{equation}
where 
\be
X  \equiv -\frac{(\partial\phi)^2}{2m^4}
\ee
is dimensionless. Here $m$ and $M$ are some mass scales to be determined by the fundamental theory. As successful merger of ekpyrosis and ghost condensation, as we will see, requires
a large hierarchy: $M \gg m$. 

The ghost condensate relies on the kinetic function $P(X)$ having a minimum at some finite $X$, taken to be $1/2$ without loss of generality, corresponding to $\dot{\phi} = -m^2$. In the absence of a potential, $X=1/2$ is an exact solution to the equations of motion, even in a cosmological background. 

\begin{figure}[ht]
\centering
\includegraphics[width=100mm]{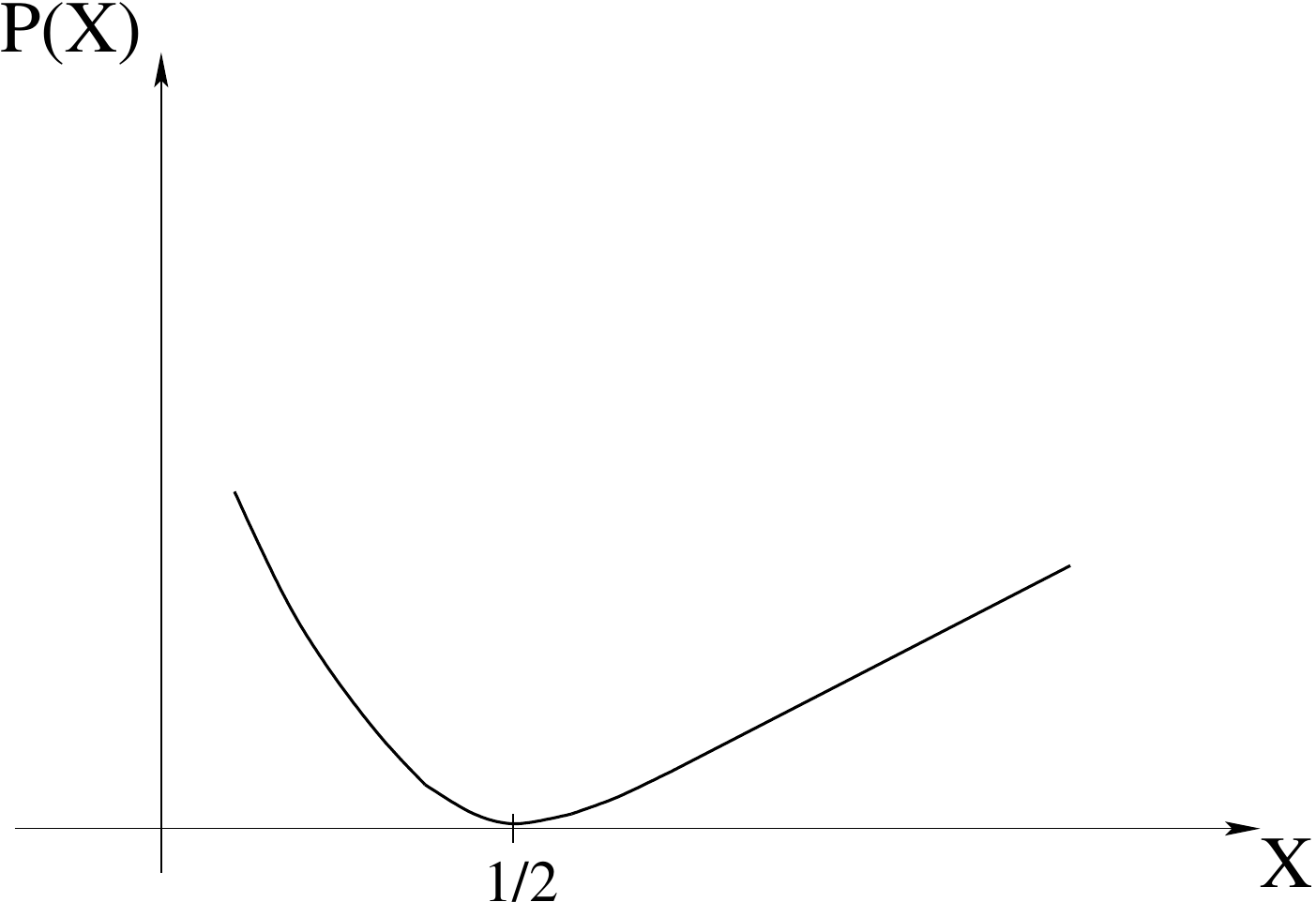}
\caption{Kinetic function for $\phi$.}
\label{P(X)}
\end{figure}

What about the global form for $P$? During the ekpyrotic phase, $\phi$ is assumed to have approximately standard kinetic term, corresponding to $P(X)\approx X$. In order for fluctuations to have positive norm, this quasi-linear part must lie to the right of the minimum, {\it i.e.} for $X\;\gsim\; 1/2$~\cite{us}. If the quasi-linear part lies at $X\;\lsim\; 1/2$, on the other hand, the field evolution must go through a region with $P_{,X}<0$, corresponding to ghost-like fluctuations~\cite{us}. The desired form of $P(X)$ is sketched in Fig.~\ref{P(X)}. 

In the process of climbing up the sharp rise in the potential, 
the kinetic energy in $\phi$ decreases dramatically, and
$X$ is brought to the vicinity of the minimum of $P(X)$ by the 
time $\phi$ reaches the plateau. This marks the onset of the ghost 
condensate phase which for convenience we will set at $t=0$.
Then, near the ghost
condensate point we can expand the field as $\phi = -m^2t + \pi$, 
leading to the following expressions for the energy density and pressure:
\bea \nonumber
\rho &=& M^4\left(2P_{,\,X}X-P\right) +V  \approx  -\frac{M^4 K \dot{\pi}}{m^2}+V \\
{\cal P} &=& M^4P(X) -V \approx\ -V\,, \label{5.31} \eea
where $K=P_{, XX}$. See~\cite{us} for details.
(Notice that we have taken $P(1/2) = 0$ without loss of generality 
since an overall shift in $P$ is degenerate with a shift in $V$.) 

The culprit in violating the NEC is the term linear in $\dot{\pi}$, since its contribution to the energy density does not have a definite sign.
Put another way, the $\dot{H}$ equation is given by
\be
M_{\rm Pl}^2\dot{H} = -\frac{1}{2}\left(\rho+{\cal P}\right) \approx \frac{M^4 K\dot{\pi}}{2m^2}\,,
\label{dotHpi}
\ee
which clearly can take either sign depending on the sign of $\dot{\pi}$. 

\vspace{0.5cm}
\noindent {\it iii) Constraints for successful bounce}
\vspace{0.2cm}

The successful merging of ekpyrosis with ghost condensation imposes some constraints on the ekpyrotic potential. 

Firstly, by assumption we must have $X\;\gsim\; 1/2$ throughout the ekpyrotic phase, that is, $\dot{\phi}^2 \gg m^4$. Because the kinetic energy grows during the ekpyrotic phase, evidently 
this condition is most stringent at the onset of this phase: $\dot{\phi}^2(t_{\rm ek-beg})\gg m^4$.
And since $\dot{\phi}^2/2\approx -V$ during the ekpyrotic phase, this immediately implies: $|V(\phi_{\rm ek-beg})| \gg m^4$. For our purposes we find it convenient to express this as
a constraint on the value of the potential at the minimum, denoted by $V_{\rm min}\equiv V(\phi_{\rm min},\chi_{\min})$. From the scaling solution~(\ref{5.2}) and~(\ref{Nek}), we have $|V(\phi_{\rm ek-beg})|  \sim e^{-2N_{\rm ek}} |V_{\rm min}|$. Therefore we must require
\be
\left\vert V_{\rm min}\right\vert \gg e^{2N_{\rm ek}}m^4\,.
\ee

Secondly, by assumption $\phi$ must lie in the vicinity of the ghost condensate point by the time it reaches the plateau, so that we can approximate $\phi\approx -m^2t + \pi$. As shown in~\cite{us}, this gives an upper bound on $V_{\rm min}$, the value of the potential at the minimum: 
\be
\left\vert V_{\rm min}\right\vert \ll \frac{M^4 K}{p}\,.
\ee

To summarize, the ekpyrotic field is successfully morphed into an NEC-violating ghost condensate provided $|V_{\rm min}|$ falls within the range 
\be
e^{2N_{\rm ek}}m^4 \ll \left\vert V_{\rm min}\right\vert \ll \frac{M^4 K}{p}\,.
\ee
Evidently this is satisfied provided $M$ is exponentially larger than $m$. For instance, for a reheating temperature of $10^{15}$~GeV corresponding to $N_{\rm ek} \;\gsim\; 60$ and $p=10^{-2}$, we can take $M\;\gsim\; 10^{16}$~GeV and $m\;\gsim\; 10^3$~GeV.

So far everything we have said is a review of aspects of ghost condensate/ekpyrosis merger first discussed in~\cite{us}. A key difference with the two-field scenario of~\cite{us}, however,
is that here we shall use only one field, namely $\phi$, as ghost condensate, while $\chi$ is stabilized. While this greatly simplifies the picture, there is an additional constraint to consider,
having to do with the energy density in the $\chi$-oscillations, denoted earlier by $\rho_\chi$. This is {\it a priori} non-trivial since the latter blueshifts as $1/a^3$ as the universe contracts
--- see~(\ref{rhoa3}) ---, whereas  the ghost energy density {\it decreases} since it violates the NEC. Nevertheless, this is not a big concern for two reasons. As mentioned earlier, a few more
e-folds of $\phi$ rolling down its steep exponential potential after the end of the ekpyrotic phase makes $\rho_\chi$ negligibly small compared to $\rho_\phi$ by the onset of the
NEC-violating era. Moreover, from the moment $\phi$ enters the ghost condensate phase until reheating, the scale factor $a(t)$ changes by at most a factor of order unity. Thus $\rho_\chi$
does not blueshift appreciably throughout the NEC-violating epoch.

\begin{figure}[ht]
\centering
\includegraphics[width=100mm]{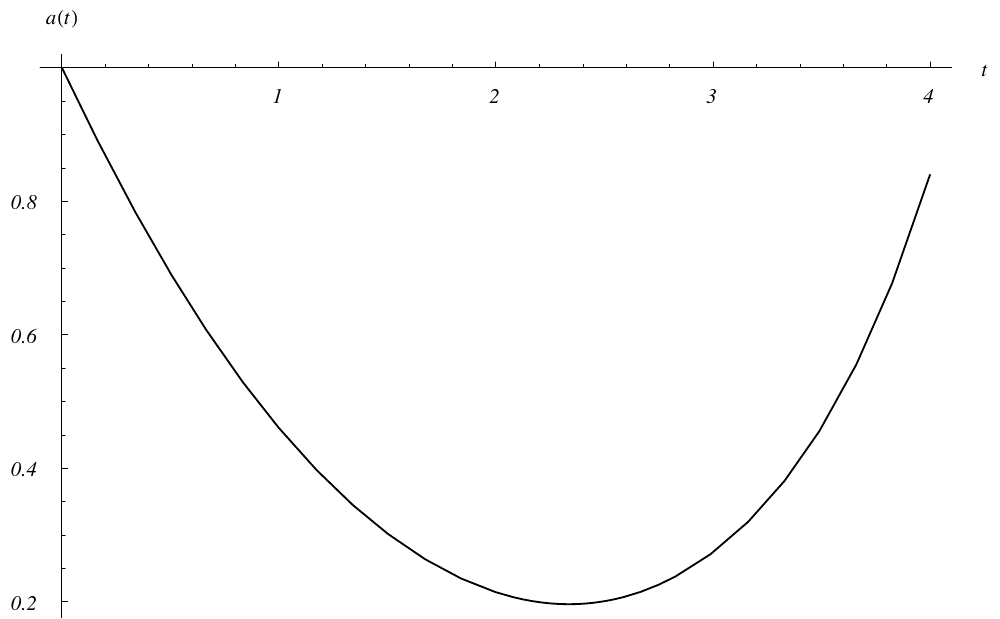}
\caption{Evolution of the scale factor $a(t)$ during the bounce generated by the ghost condensate. The NEC-violating phase begins by definition when $\dot{H}=0$, which as shown
in Fig.~\ref{H(t)} occurs around $t\approx 1.2$ in these units. Meanwhile the bounce occurs by definition when $\dot{a} = 0$, which occurs around $t\approx 2.4$. The point is that in this time
interval the scale factor is seen to change by a factor of 2, from approximately 0.4 to 0.2.
Note that we have set $t=0$ to be the beginning of the ghost condensation phase.}
\label{a(t)}
\end{figure}

\begin{figure}[ht]
\centering
\includegraphics[width=100mm]{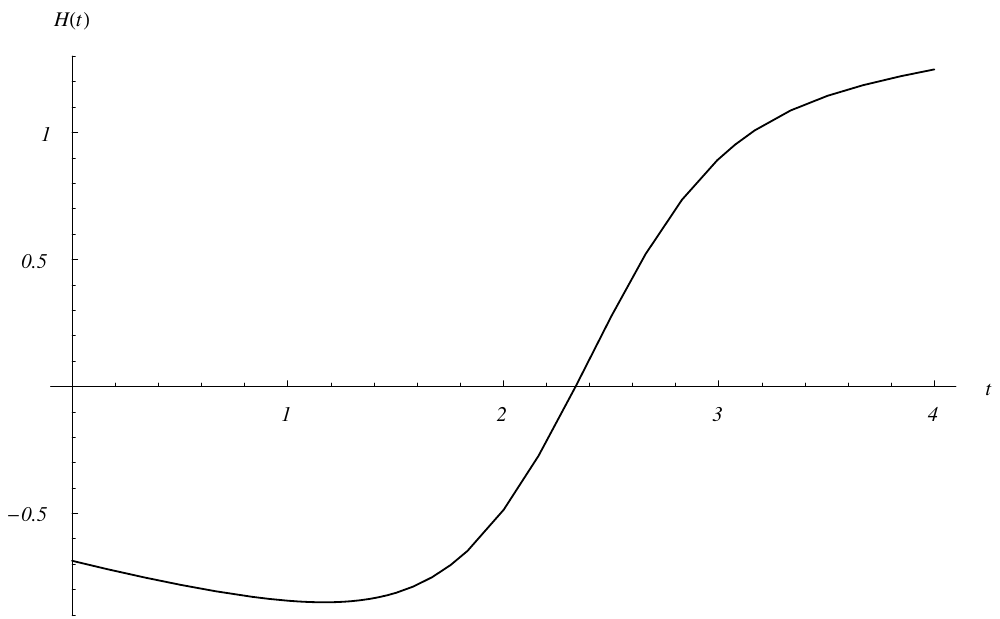}
\caption{Evolution of the Hubble parameter $H(t)$ during the bounce generated by the ghost condensate. The epoch of NEC violation starts when $\dot{H} = 0$, which from the plot is seen to occur at
$t\approx 1.2$. The bounce ($H=0$) occurs at $t\approx 2.4$. Note that we have set $t=0$ to 
be the beginning of the ghost condensation phase.}
\label{H(t)}
\end{figure}

It remains to show that $a(t)$ satisfies this property. Clearly this is the case from the end of ekpyrosis until $\phi$ reaches the plateau, since everything happens almost instantaneously on a Hubble time, by assumption. 
That this is also satisfied throughout the subsequent NEC-violating phase follows from requiring that the bounce occurs within at most a Hubble time or so, which, as argued in~\cite{Paolo}, 
ensures that the Jeans-like instabilities intrinsic to ghost condensation remain under control during the bounce. A numerical solution obtained in~\cite{us} and plotted in Figs.~\ref{a(t)} and~\ref{H(t)} 
shows that the scale factor changes by approximately a factor of 2 over the course of the NEC-violating phase. See the captions for details.
Consequently, $\rho_\chi$ blueshifts by a factor of 8 during this phase.

Meanwhile, it is easily seen that the ghost condensate energy density remains essentially constant during this phase. To estimate its value, we assume for simplicity that the
scaling solution~(\ref{5.2}) holds all the way to the onset of the ghost condensate phase. In this case we have
\be
\rho_{\rm ghost}\approx  p\left\vert V_{\rm min}\right\vert\,.
\ee
Therefore, ignoring the effect of the ekpyrotic attractor mechanism between $t_{\rm ek-end}$ until $t_c$, a sufficient condition to ensure the subdominance of $\rho_\chi$ during the bounce is
\be
\rho_\chi(t_{\rm ek-end}) = V\left(\phi_{\rm ek-end}, 0\right) - V\left(\phi_{\rm ek-end},\chi_{\rm min}\right)\;\lsim \; 8p\left\vert V_{\rm min}\right\vert \,,
\label{rhorad2}
\ee
where the factor of 8 comes from the blueshift of $\rho_\chi$ through the bounce, as described above.
In other words, the potential drop experienced by $\chi$ as it rolls towards its stable minimum must be small compared to $V_{\rm min}$.

\subsection{Summary}
\label{summary}

Let us summarize the sequence of events from the end of the ekpyrotic phase through the non-singular bounce and onto the radiation-dominated hot big bang phase:

\begin{itemize}

\item The entropy field $\chi$ is eventually driven away from the tachyonic ridge, thereby bringing to a halt the production of scale-invariant density perturbations. See Fig.~\ref{2fieldpotend}. This is achieved by a correction to the ekpyrotic potential~(\ref{4.7}), for instance of the form~(\ref{egpot}), which has non-zero slope at $\chi=0$. For simplicity, we assume that the departure from $\chi\approx 0$ occurs rapidly on a Hubble time, which requires a fast-roll condition to be satisfied
\be
|V_{,\chi}(\chi=0)| \;\gsim\; H^2M_{\rm Pl}\,.
\label{sumcond1}
\ee
We denote by $\phi_{\rm ek-end}$ the value of $\phi$ at which $|V_{,\chi}|$ starts being comparable to $H^2M_{\rm Pl}$.

\item Subsequently $\chi$ rolls to a stable minimum denoted by $\chi_{\rm min}$ and undergoes oscillations. As $\chi$ acquires a non-zero expectation value, the light fermions
of the pre-ekpyrotic phase become heavy and cannot be relied upon to stabilize $\chi$. We also impose that $\chi$ is heavy at its minimum,
\be
V_{,\chi\chi}\left(\chi_{\rm min}\right)\gg H^2\,,
\label{largemass}
\ee
so that its oscillations are rapid on a Hubble time. The energy density acquired by $\chi$, given by $\rho_\chi(t_{\rm ek-end}) =  V\left(\phi_{\rm end}, 0\right) - V\left(\phi_{\rm end},\chi_{\rm min}\right)$,
blueshifts effectively as dust and quickly becomes subdominant to $\phi$.

\item The evolution then reduces to single-field ekpyrosis. The end of the ekpyrotic phase is triggered by $\phi$ reaching a minimum in the potential,
which subsequently rises to positive values where it becomes flat. We denote by $V_{\rm min}$ the value of the potential at the minimum, and by $\phi_{\rm min}$ the corresponding field value.
Since stabilization of $\chi$ is assumed to occur before $\phi$ hits the sharp rise, for consistency we must have
\be
\phi_{\rm min} \;\lsim\; \phi_{\rm ek-end}\,.
\ee

\item As $\phi$ climbs up the sharp rise of the potential, its kinetic energy decreases substantially, which brings $X$ to the vicinity of the ghost condensate point. This marks the beginning of the NEC-violating phase. A successful synergy of ekpyrotic and ghost condensate cosmology requires $V_{\rm min}$ to fall within the allowed window
\be
e^{2N_{\rm ek}}m^4 \ll \left\vert V_{\rm min}\right\vert \ll \frac{M^4 K}{p}\,,
\ee
where $m$ and $M$ parametrize the location of the minimum and curvature of the kinetic function $P(X)$, respectively. 

\item In order to have a bounce, the energy density in the ghost condensate, given by $pV_{\rm min}$, must dominate over the energy density in $\chi$ oscillations, $\rho_\chi$.
Since the latter blueshifts by a factor of 8 or so during the bounce, this leads to the constraint:
\be
V\left(\phi_{\rm ek-end}, 0\right) - V\left(\phi_{\rm ek-end},\chi_{\rm min}\right)\;\lsim \; 8p\left\vert V_{\rm min}\right\vert \,.
\label{rhoradcond}
\ee

\end{itemize}

These five equations are the main constraints to have a successful exit from ekpyrotic phase followed by a non-singular bounce. In the next section, we illustrate how these conditions can be fulfilled by studying a specific potential consisting of the ekpyrotic potential~(\ref{4.7}) with the fiducial correction term~(\ref{egpot}).


\section{A Specific Exit Model}
\label{example}


We now illustrate the constraints described in the previous section with an explicit choice of potential. Our fiducial potential consists of two
terms: $i$) the ekpyrotic potential~(\ref{Vbew}) which we assume has exact exponential form in $\phi$; $ii$) the correction term~(\ref{egpot}) which generates
a minimum in $\phi$ and forces $\chi$ away from the tachyonic ridge. Thus the full potential is given by
\be
V(\phi,\chi) =  V_{\rm ek}(\phi,\chi) + \Delta V(\phi,\chi)\,,
\ee
with
\begin{eqnarray}
\nonumber
V_{\rm ek}(\phi,\chi) &=& - V_0
e^{-\sqrt{\frac{2}{p}}\phi/M_{\rm{Pl}}} \left(1+\frac{1}{p
M^2_{\rm{Pl}}}\chi^2
+\dots\right) \\
 \Delta V(\phi,\chi) & = & U_0e^{-\sqrt{\frac{2}{p^{\prime}}}\phi/M_{\rm{Pl}}}
\left(1+\frac{1}{p^{\prime}M^2_{\rm{Pl}}}(\chi-\chi_0)^2 +\dots\right)\,,
\label{5.9}
\end{eqnarray}
where $V_0$ and $U_0$ are both positive, and, as earlier, ellipses denote higher-order terms in $\chi$ which are irrelevant for our purposes. Note that we neglect the mass term for $\chi$ added in~(\ref{4.7}). While this term is crucial in stabilizing $\chi$ early on, it becomes subdominant by the onset of the ekpyrotic phase and remains subdominant
forever after. The first term $V_{\rm ek}$ is the most constrained part of the potential since it pertains to the generation of density perturbations, whereas there is considerable freedom in the correction term $ \Delta V$. We focus here on this simple form for the latter given its manifest resemblance to the ekpyrotic piece.

The exponents satisfy $p'\ll p$, which guarantees that $\Delta V$ is negligible during the ekpyrotic phase where $\phi$ is large. As $\phi$ decreases, however,
eventually this term becomes relevant and pushes $\chi$ away from the tachyonic point, triggering the end of ekpyrosis.

\subsection{Global Minimum for $V$}

This potential has a global minimum. Under some mild assumptions we can
derive explicit analytical expressions for the corresponding field values $\phi_{\rm min}$ and $\chi_{\rm min}$. 
First note that the condition $\partial V/\partial\phi = 0$ can be made nearly independent of $\chi$ if we assume
that the $\cO(\chi^2)$ terms are small compared to unity, at least in the relevant range $0\;\lsim\;\chi\;\lsim\;\chi_0$.
(We will check {\it a posteriori} that $\chi_{\rm min}\approx \chi_0$ so that this is indeed the range of interest.) 
Since $p'\ll p$ it is sufficient to impose
\begin{equation}
\frac{\chi_0}{M_{\rm{Pl}}}\ll \sqrt{p'} \,. \label{5.14}
\end{equation}
And since $p'\ll p\ll 1$ this says that the minimum along $\chi$ from the $\Delta V$ term lies at small values compared to $M_{\rm Pl}$. In other words, $\chi$ moves a small distance in field space as it rolls away from the tachyonic point towards the minimum. We stress that this condition is not necessary but is only imposed to simplify the expression for $\phi_{\rm min}$. We find
\begin{equation}
\phi_{\rm min} = \frac{M_{\rm Pl}}{\sqrt{2}}\frac{\sqrt{pp'}}{\sqrt{p}-\sqrt{p'}}
\ln \left( \sqrt{\frac{p}{p^{\prime}}}
\frac{U_0}{V_0}\right)\,. \label{5.15}
\end{equation}

Next we can solve $\partial V/\partial\chi =0$ for $\chi_{\rm min}$ by substituting the above expression for $\phi_{\rm min}$ everywhere:
\be
\chi_{\rm min} =\frac{\chi_0}{1-\frac{p^{\prime}}{p}\frac{V_0}{U_0}
e^{\left(\sqrt{\frac{2}{p^{\prime}}}-\sqrt{\frac{2}{p}}\right)\phi_{\rm min}/M_{\rm{Pl}}}} =   \frac{\chi_0}{ 1-\sqrt{\frac{p^{\prime}}{p}}}\approx \chi_0\,.
\label{5.10}
\end{equation}
It follows that $\Delta V$ plays the dominant role in determining $\chi_{\rm min}$.

\subsection{Exit from Ekpyrosis: $\chi$ Starts to Roll}

The next step is to find when $\chi$ starts to roll away from the tachyonic point at $\chi\approx 0$ towards the minimum at $\chi_{\rm min}\approx \chi_0$. 
As described in Sec.~\ref{summary}, this occurs at a value $\phi_{\rm ek-end}$ at which 
\begin{equation}
\left\vert V_{,\chi} (\chi=0)\right\vert \sim M_{\rm{Pl}}H^2\,. \label{5.19}
\end{equation}
From~(\ref{5.9}) the left-hand side is given by
\be
V_{,\chi} (\chi=0) = -U_0e^{-\sqrt{\frac{2}{p^{\prime}}}\phi_{\rm ek-end}/M_{\rm{Pl}}}\frac{2}{p^{\prime}}\frac{\chi_0}{M_{\rm{Pl}}}\,.
\ee
Meanwhile, since the ekpyrotic phase is still going on until $\phi = \phi_{\rm ek-end}$, we can use the scaling solution~\eqref{5.2}
to determine the right-hand side:
\be
H_{\rm ek-end}^2 M^2_{\rm{Pl}}\approx p V_0
e^{-\sqrt{\frac{2}{p}}\phi_{\rm ek-end}/M_{\rm{Pl}}}\,. \label{5.21}
\end{equation}
Combining the above two expressions allows us to solve for $\phi_{\rm end}$:
\be
\phi_{\rm ek-end}=\frac{M_{\rm Pl}}{\sqrt{2}}\frac{\sqrt{pp'}}{\sqrt{p}-\sqrt{p'}}
\ln \left( \frac{2}{p p^{\prime}} \frac{U_0}{V_0}
\frac{\chi_0}{M_{\rm{Pl}}}\right)\,. \label{5.22}
\end{equation}

To summarize, for $\phi\;\gsim\; \phi_{\rm end}$, a good approximate solution for $\chi$ is just $\chi\approx 0$. This is the ekpyrotic phase, characterized
by rolling in the $\phi$ direction while $\chi$ sits idle at the top of the tachyonic ridge. Once $\phi$ reaches $\phi_{\rm end}$, however, we have $|V_{,\chi}|\sim H^2M_{\rm Pl}$ and
$\chi$ is pushed away from its unstable point. Subsequently we have $|V_{,\chi}|\; \gsim\; H^2M_{\rm Pl}$, thereby satisfying the first condition~(\ref{sumcond1}) of Sec.~\ref{summary}.

\subsection{Mass of $\chi$ about Minimum}

Once $\chi$ reaches $\chi_{\rm min}\approx \chi_0$, it undergoes oscillations about this stable point.The second condition~(\ref{largemass}) ensures that the mass of
$\chi$ around $\chi_{\rm min}\approx \chi_0$ is large compared to Hubble, so that the field undergoes many oscillations in a Hubble time.
For simplicity we assume that $\phi$ is nearly still as $\chi$ rolls to the minimum, so that all relevant quantities can be evaluated at $\phi_{\rm ek-end}$.

Using~\eqref{5.22} we obtain
\begin{equation}
V_{,\chi\chi}\left(\phi_{\rm min},\chi_{\rm min}\right)  =\frac{pV_0}{M^2_{\rm{Pl}}}
e^{-\sqrt{\frac{2}{p^{\prime}}}\phi_{\rm ek-end}/M_{\rm{Pl}}}
\left(\frac{M_{\rm{Pl}}}{\chi_0}-\frac{2}{p^2}\right)\,. \label{5.26}
\end{equation}
Evidently this must be positive if we want $\chi_{\rm min}$ to be a stable point by the time we reach $\phi_{\rm ek-end}$. The condition $V_{,\chi\chi}(\phi_{\rm min})> 0$ amounts to
\begin{equation}
\frac{\chi_0}{M_{\rm{Pl}}}< \frac{p^2}{2}\,.
\label{5.26.1}
\end{equation}
Much like~\eqref{5.14}, this forces $\chi_0$ to be small in Planck units. 

Using~(\ref{5.21}) we obtain
\begin{equation}
\frac{V_{,\chi\chi}\left(\phi_{\rm min},\chi_{\rm min}\right)}{H^2_{\rm ek-end}}=\frac{M_{\rm{Pl}}}{\chi_0}-\frac{2}{p^2}\,.
\label{5.28}
\end{equation}
Barring some fine-tuning between these two terms, this ratio is generically much bigger than unity. 
It then follows that $\chi$ acquires a large mass at the minimum and oscillates rapidly.

\subsection{Exit Happens Before $\phi$ Reaches Minimum}

Let us turn to the third condition, which requires that $\phi_{\rm min}\;\lsim \; \phi_{\rm ek-end}$. This ensures consistency of the sequence of events assumed here, namely that
$\chi$ starts to roll towards its minimum, thereby imprinting the entropy perturbation onto $\zeta$, before $\phi$ reaches the steep rise in the potential. Again this assumption is not necessary ---
nothing prevents us from converting the entropy mode into the curvature perturbation after the $\phi$ has reached its minimum. However an estimate of $\zeta$ would probably require numerical analysis in this case. In any case, comparison of~(\ref{5.22}) with~(\ref{5.15}) shows that $\phi_{\rm min}\;\lsim \; \phi_{\rm end}$ is satisfied if
\be
\frac{\chi_0}{M_{\rm{Pl}}} \;\gsim\; p^{3/2} \sqrt{p^{\prime}}\,.
\label{5.24}
\end{equation}
Note that this condition is consistent with~\eqref{5.14} and~\eqref{5.26.1} since $p'\ll p \ll 1$. 

\subsection{Energy Density in $\chi$ Oscillations}

The last condition~(\ref{rhoradcond}) puts an upper bound on the energy density in $\chi$ as it rolls to the minimum and starts oscillating.
As argued in Sec.~\ref{end}, this is just given by the difference in potential energy between the ridge and the minimum --- see~\eqref{rhorad}.
Since $V$ contains exponential factors, evidently this difference is maximal in the limit $\phi_{\rm min}\rightarrow \phi_{\rm ek-end}$.  Thus we find
\begin{eqnarray}
\Delta V  & < & V_0
e^{-\sqrt{\frac{2}{p}}\phi_{\rm min}/M_{\rm{Pl}}}\frac{\chi_0^2}{p
M^2_{\rm{Pl}}}\left(1+ \sqrt{\frac{p}{p^{\prime}}}\right) \nonumber \\
\nonumber
& \approx & V_0 e^{-\sqrt{\frac{2}{p}}\phi_{\rm min}/M_{\rm{Pl}}}
\frac{\chi_0^2}{\sqrt{p p^{\prime}} M^2_{\rm{Pl}}}\,.
 \label{5.40}
\end{eqnarray}
Meanwhile $V_{\rm min}$ is to a good approximation given by $V_{\rm ek}$:
\be
|V_{\rm min}| \approx V_0e^{-\sqrt{\frac{2}{p}}\phi_{\rm min}/M_{\rm{Pl}}}\,.
\ee

Now from~(\ref{5.15}) and~(\ref{5.22}) we see that the limiting case $\phi_{\rm min} = \phi_{\rm ek-end}$ is achieved for $\chi_0 = p^{3/2}\sqrt{p'}M_{\rm Pl}/2$. Thus the relevant ratio for condition~(\ref{rhoradcond}) can be written as
\be
\frac{\Delta V}{p|V_{\rm min}|} \approx \frac{\chi_0}{2 M_{\rm{Pl}}}\,,
\label{5.44}
\ee
which has to be much less than unity. But this is clearly the case since~(\ref{5.14}) implies $\chi_0\ll M_{\rm Pl}$.
It follows that the energy density in $\chi$ oscillations is parametrically small compared to the ghost condensate energy density,
thereby ensuring that the NEC is violated for sufficiently long to cause a non-singular bounce. \\

\section{Conclusion}
\label{conclude}

The many aspects of ekpyrotic cosmology discussed in the paper all revolve around the issue of initial conditions.
Unlike the hot and chaotic beginning of inflation, ekpyrosis proposes a cold and nearly vacuous initial state. Put more
concretely, for GUT-scale reheating temperature, the proto-inflationary patch is $10^{-27}$~cm big, whereas the
initial Hubble radius in ekpyrosis is of order 1~m. In the absence of a concrete theory of initial conditions, however,
both represent equally plausible initial states.

We have shown that ekpyrotic theory is equally successful as inflation in solving the flatness and homogeneity problems of
standard big bang cosmology. Indeed, the model allows for initial curvature and anisotropic components comparable to that of the
ekpyrotic field. Starting from these generic initial conditions, the universe emerges in the hot, expanding phase with a high degree of spatial flatness,
homogeneity and isotropy. Note that this is very similar to how inflation addresses these problems --- one envisions a
universe that is essentially smooth and flat over the proto-inflationary patch, thereby allowing cosmic acceleration to take over. 

The two-field ekpyrotic model displays a tachyonic instability along $\chi$, the direction orthogonal to the field trajectory. 
Since fluctuations in $\chi$ by definition coincide with entropy perturbations, this instability is essential in generating a scale-invariant
spectrum. We cannot get rid of it without spoiling the perturbation spectrum. We have shown that the desired initial conditions at the onset
of the ekpyrotic phase can be naturally achieved with a pre-ekpyrotic stabilizing phase. By adding a small mass for $\chi$ and couplings
to light fermions, $\chi$ gets stabilized well-before the onset of the ekpyrotic phase for a broad range of initial conditions. The mass term is
only important at early times --- it becomes negligibly small during the ekpyrotic phase and therefore does not interfere with the generation of
density perturbations. 

The analysis of the tachyonic instability in terms of new field variables $\phi$ and $\chi$ lead us to propose a simplified and more general version
of the scenario. Instead of starting from two scalar fields each with their own exponential potential, we can think of one field $\phi$ rolling down an ekpyrotic direction whose evolution determines the tachyonic mass of a second field $\chi$. This occurs in such a way that fluctuations in the latter acquire a scale-invariant spectrum. 
The ($\phi,\chi$) language allows for far greater freedom in specifying the potential. Whereas the earlier $(\phi_1,\phi_2)$ framework required two steep exponential functions in the potential,
here we only need one steep potential along $\phi$ and one coefficient in a Taylor-expansion --- the mass term for $\chi$ --- to be such that the curvature of the potential is approximately the same in either direction: $V_{,\chi\chi} \approx V_{,\phi\phi}$. 
There is, therefore, enormous freedom in specifying the global shape of the potential.

The simplified picture in terms of $\phi$ and $\chi$ has important consequences for the predictions of the model,
in particular for the spectral index and the level of non-Gaussianity. The former now depends on an additional
parameter $\delta$ that characterizes the deviation from $V_{,\chi\chi} = V_{,\phi\phi}$. With vanishing $\delta$, 
a pure exponential for $\phi$ leads to a slightly blue spectrum, in disagreement with recent data. Allowing for
non-zero $\delta$, however, can make the spectral tilt red even in this case. Turning to non-Gaussianity, we have argued that self-interactions in $\chi$ have a natural cut-off scale of
order $\Lambda = \sqrt{\epsilon}M_{\rm Pl}$. Assuming coefficients of order unity, the resulting level of
non-Gaussianity is therefore high, corresponding to $f_{\rm NL}\sim 1/\epsilon$. The consequences for current and near-future
microwave background experiments will be discussed elsewhere.

These achievements, we believe, together put the New Ekpyrotic scenario on a stronger footing as a genuine alternative theory of early universe cosmology.

\bigskip
{\bf Acknowledgments}
We are most grateful to Andrea Sweet at PI for invaluable help with the figures. We thank K.~Koyama, P.~Steinhardt, A.~Tolley and N.~Turok for insightful discussions.
This work of B.A.O. is supported in part by the DOE under contract No. DE-AC02-76-ER-03071 and by the NSF Focused Research Grant DMS0139799.
The research of E.I.B. and J.K. at Perimeter Institute is supported in part by the Government of Canada through NSERC and by the Province of Ontario through MRI. \\


\end{document}